\documentclass[reprint,amsmath,amssymb,aps,prr,superscriptaddress]{revtex4-2} 

\usepackage{graphicx}
\usepackage{dcolumn}
\usepackage{bm}
\usepackage{bbold}
\usepackage[pdfusetitle,
            bookmarks=true,
            colorlinks,
            linkcolor=blue,
            citecolor=blue,
            urlcolor=blue
            ]{hyperref}
\usepackage{xcolor}
\usepackage{gensymb}
\usepackage{listings} 

\usepackage{physics}

\definecolor{codegreen}{rgb}{0,0.6,0}
\definecolor{codegray}{rgb}{0.5,0.5,0.5}
\definecolor{codepurple}{rgb}{0.58,0,0.82}
\definecolor{backcolour}{rgb}{0.95,0.95,0.92}

\lstdefinestyle{mystyle}{
    backgroundcolor=\color{backcolour},   
    commentstyle=\color{codegreen},
    keywordstyle=\color{magenta},
    numberstyle=\tiny\color{codegray},
    stringstyle=\color{codepurple},
    basicstyle=\ttfamily\footnotesize,
    breakatwhitespace=false,         
    breaklines=true,                 
    captionpos=b,                    
    keepspaces=true,                 
    numbers=left,                    
    numbersep=5pt,                  
    showspaces=false,                
    showstringspaces=false,
    showtabs=false,                  
    tabsize=2
}

\newcommand\numberthis{\addtocounter{equation}{1}\tag{\theequation}}

\newcommand{\eps}{\varepsilon}

\newcommand{\kk}{\mathbf{k}}

\newcommand{\pbold}{\mathbf{p}}

\newcommand{\qb}{\mathbf{q}}

\newcommand{\mfp}{\chi}

\newcommand{\bexpval}[1]{\big\langle#1\big\rangle}
\newcommand{\vbar}{\mathcal{V}}
\newcommand{\nbar}{\theta}

\begin{document}

\title{Purely Electronic Model for Exciton-Polaron Formation in Moir\'e Heterostructures}

\newcommand{\TUM}{\affiliation{Technical University of Munich, TUM School of Natural Sciences, Physics Department, 85748 Garching, Germany}}
\newcommand{\MCQST}{\affiliation{Munich Center for Quantum Science and Technology (MCQST), Schellingstr. 4, 80799 M{\"u}nchen, Germany}}
\newcommand{\JQI}{\affiliation{Joint Quantum Institute, University of Maryland, College Park, Maryland 20742, USA}}

\author{Fabian Pichler} \TUM \MCQST
\author{Mohammad Hafezi} \JQI
\author{Michael Knap} \TUM \MCQST

\date{\today}

\begin{abstract}
Understanding interactions between excitons and correlated electronic states presents a fundamental challenge in quantum many-body physics. Here, we introduce a purely electronic model for the formation of exciton-polarons in moir\'e lattices. Unlike conventional approaches that treat excitons as tightly-bound bosonic particles, our model considers only electronic degrees of freedom, describing excitons as electron-hole bound states. Our findings reveal a pronounced renormalization of the polaron mass as a function of electron density, particularly near correlated insulators, consistent with recent transport experiments. Additionally, we predict an observable sign change in the effective polaron mass when increasing the electron density that can be measured in Hall-type experiments. Our purely electronic model provides a unified framework to investigate the formation and renormalization of exciton-polarons in correlated states. 
\end{abstract}

\maketitle

\section{Introduction}

Excitons have proven to be a valuable tool for probing strongly-correlated states in transition metal dichalcogenide (TMD) heterostructures, including periodic charge modulation~\cite{regan_mott_2020, shimazakiStronglyCorrelatedElectrons2020, xu_correlated_2020, shimazakiOpticalSignaturesCharge2021, Smolenski_ObservationWigner_2021}, spin ordering ~\cite{ciorciaroKineticMagnetismTriangular2023, tang_evidence_2023}, and dipolar excitonic insulators~\cite{guDipolarExcitonicInsulator2022a, zhangCorrelatedInterlayerExciton2022, mhenniGatetunableBoseFermiMixture2024}.
An increasing number of experiments focus on the physics of excitons in moir\'e heterostructures~\cite{riveraObservationLonglivedInterlayer2015,  tranEvidenceMoireExcitons2019, seylerSignaturesMoiretrappedValley2019, jinObservationMoireExcitons2019, baekHighlyEnergytunableQuantum2020, miaoStrongInteractionInterlayer2021a, wangMoireTrionsMoSe22021,parkDipoleLaddersLarge2023,kaiDistinctMoireExciton2024,kiperConfinedTrionsMottWigner2024, devenica2025}, with a particular focus on transport and diffusion experiments~\cite{choiMoirePotentialImpedes2020,
yuanTwistangledependentInterlayerExciton2020,
malicExcitonTransportAtomically2023, rossiAnomalousInterlayerExciton2024,  upadhyayGiantEnhancementExciton2024, yanAnomalouslyEnhancedDiffusivity2024}, yet our theoretical understanding of exciton dynamics in lattice systems and strongly correlated environments remains vastly open. When immersed in a strongly correlated bath, excitons become dressed by their environment, forming a quasiparticle known as polaron. While some recent studies have highlighted key aspects of exciton-polaron formation in moir\'e lattices~\cite{mazzaStronglyCorrelatedExcitonpolarons2022, knorrExcitonTransportMoire2022, gottingMoireBoseHubbardModelInterlayer2022, huangMottmoireExcitons2023, huangNonbosonicMoireExcitons2024, knorrPolaroninducedChangesMoire2024, julkuExcitonInteractingMoire2024, huangCollectiveOpticalProperties2024, zhengForsterValleyorbitCoupling2024, shentsevElectromagneticFieldAssisted2025, evrardAcStarkSpectroscopy2025} and other correlated systems~\cite{ yiPolaronsUltracoldFermi2015, colussiLatticePolaronsSuperfluid2023, amelioTwodimensionalPolaronSpectroscopy2023a, santiago-garciaLatticePolaronBoseEinstein2024, amelioPolaronSpectroscopyInteracting2024, amelioPolaronFormationInsulators2024}, many open questions remain.
In particular, the simplification of excitons as tightly-bound bosonic particles without internal structure is insufficient to capture the behavior of interlayer excitons in TMD heterostructures~\cite{huangNonbosonicMoireExcitons2024}. Moreover, it remains unclear which role different stacking configurations have on the dressed exciton-polaron.
Developing effective theoretical models suited to describe the exciton-polaron dynamics in moir\'e lattices is therefore pertinent to help interpret already existing data and guide future experimental investigations.

\begin{figure}
\begin{center}
\includegraphics[width=0.99\linewidth]{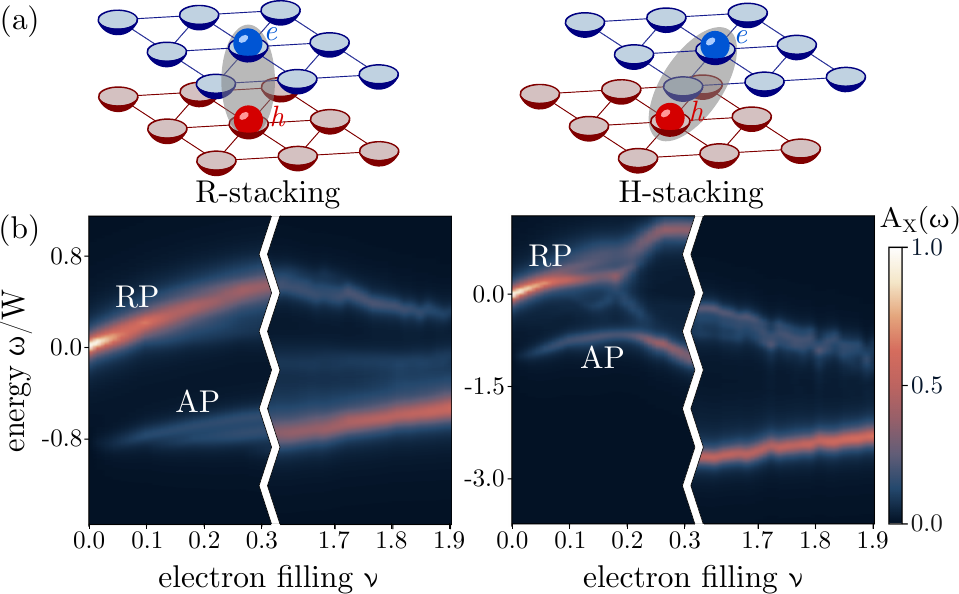}
\caption{\textbf{Exciton-polaron in low and high-density regime.} \textbf{(a)} Schematics of interlayer-exciton formation in moir\'e lattices. The top layer is populated by electrons (blue sphere), and the bottom layer by holes (red sphere). For R-stacking (left), the electron and hole sites lie on top of each other while they are shifted for H-stacking. \textbf{(b)} Zero-momentum exciton spectral function, showing the repulsive polaron (RP)  and the attractive polaron (AP), which appears at finite electron filling $\nu$. The energy is given in units of the bare electron bandwidth $W=9t$, with $t$ the electron hopping. We use the same local electron-hole interactions ($U_{e-h}=-15t$) for both stackings; all other interactions are set to zero. System size $N = 12\times 12$.}
\label{fig:1}
\end{center}
\end{figure}

In this work, we develop a simple yet powerful, purely electronic model to describe the dynamics of an interlayer-exciton in a moir\'e lattice. Our model only contains electronic degrees of freedom and fully accounts for the internal structure of the exciton as a bound state of an electron and hole. Employing the Chevy approximation~\cite{chevyUniversalPhaseDiagram2006}, we calculate the exciton propagator in the presence of a Fermi liquid and correlated insulators. At low electron densities, the model reproduces the characteristic formation of attractive (AP) and repulsive polaron (RP) branches, consistent with the continuum~\cite{sidlerFermiPolaronpolaritonsChargetunable2017, imamogluExcitonpolaronsTwodimensionalSemiconductors2020}. Beyond this, our model provides insight into the strong renormalization of exciton mobility as a function of electron filling, particularly near Mott insulating states, as observed in recent diffusion experiments~\cite{upadhyayGiantEnhancementExciton2024, yanAnomalouslyEnhancedDiffusivity2024}. We predict a sign reversal of the effective polaron mass at a critical filling and propose a Hall-type measurement for its experimental verification. Our purely electronic model, therefore, provides a unified framework for the exciton-polaron formation in correlated states that is beyond the reach of conventional Bose-Fermi models.

{

\section{Model} \label{sec:model}
    Heterostructures of TMDs have been shown to host long-lived interlayer excitons where the electron and hole reside in different layers~\cite{riveraObservationLonglivedInterlayer2015, riveraInterlayerValleyExcitons2018, torunInterlayerIntralayerExcitons2018}. To describe the dressing of excitons in correlated environments, we propose a purely electronic model that incorporates the internal structure of an exciton as a bound state of an electron and a hole. Specifically, we consider a heterobilayer system in which, at charge neutrality, the moiré lattice of the bottom layer is fully filled while the top layer is empty~\cite{wangMoireTrionsMoSe22021, liuSignaturesMoireTrions2021, wangIntercellMoireExciton2023}. In the deep-moiré regime, each layer is modeled as a triangular lattice, with holes $h^\dagger$ in the bottom layer and electrons $c^\dagger$ in the top layer. The Hamiltonian of the system is
    \begin{equation}
        H = H_e + H_h + H_{e-h}.\label{eq:Hamiltonian}
    \end{equation}
    The electrons in the top layer are governed by
    \begin{equation}
        H_e  = \displaystyle\sum_{\kk\sigma}\eps^e_\kk c^\dagger_{\sigma \kk} c_{\sigma \kk} + \displaystyle\sum_{\substack{\kk \kk' \qb \\ \sigma \sigma'}}V^{e-e}_\qb c^\dagger_{\sigma, \kk + \qb} c^\dagger_{\sigma', \kk'-\qb} c_{\sigma'\kk'}c_{\sigma \kk}, \label{eq:Ham_elec}
    \end{equation}
    including a repulsive electron-electron interaction $V^{e-e}_\qb$. 
    We assume electron doping and work in the exciton impurity limit, corresponding to a low pump intensity in experiments, see e.g. Ref.~\cite{upadhyayGiantEnhancementExciton2024}. Thus, there is only a single hole in the system, such that for the hole we only need to consider the kinetic term:
    \begin{equation}
        H_h = \displaystyle\sum_{\kk\sigma}\eps^h_\kk h^\dagger_{\sigma \kk} h_{\sigma \kk}, \label{eq:Ham_hole}
    \end{equation}
    where $\eps^{e/h}_\kk$ is the dispersion of a particle hopping on a triangular lattice. We assume the nearest-neighbor hopping as $t_e = t_h \equiv t$.
    We also include attractive electron-hole interactions $V_\qb$ between the two layers, responsible for the formation of interlayer excitons:
    \begin{equation}
        H_{e-h} = \displaystyle\sum_{\kk \kk' \qb}\displaystyle\sum_{\sigma \sigma'} V_\qb c^\dagger_{\sigma, \kk + \qb} h^\dagger_{\sigma', \kk'-\qb} h_{\sigma', \kk'}c_{\sigma \kk}. \label{eq:Ham_eh}
    \end{equation}
In practice, we restrict all interactions to short-range on-site $U$, nearest-neighbor density-density $V$, and nearest-neighbor direct-exchange $X$ interactions. Electron-electron interactions are repulsive $U_{e-e}, V_{e-e} > 0$, while electron-hole interactions are attractive $U_{e-h}, V_{e-h} < 0$.
Our purely electronic model allows us to study the exciton formation in different stacking configurations of the heterobilayer. We consider both R- and H-stacking. In the former, the lattice sites for the electrons and holes are on top of each other, while in the latter, they are shifted~\cite{wangIntercellMoireExciton2023}; see Fig.~\ref{fig:1}(a). In our model, the two distinct stacking configurations manifest in different interlayer interactions $V_\qb$, given explicitly in Appendix~\ref{sec:app:attracInt}.

We work in the low exciton-density regime, where a single exciton $x^\dagger$ is added to the electronic ground state. The exciton is treated as a composite particle. Its interlayer wavefunction for total momentum $\pbold$ is  
\begin{equation}
    \ket{X_\pbold} \equiv x^\dagger_\pbold \ket{GS} = \displaystyle\sum_\kk \psi_{\kk}(\pbold) c^\dagger_{\uparrow \pbold - \kk} h^\dagger_{\uparrow \kk} \ket{GS}. \label{eq:exitonWavefunc}
\end{equation}
We determine the wavefunction $\psi_\kk(\pbold)$ by solving the two-particle Schr{\"o}dinger equation in the presence of the electron ground state $\ket{GS}$, which we for now choose to be a Fermi sea. We later relax this assumption, discussing correlated insulators as electronic ground states, which modifies the decomposition of the exciton; see Appendix~\ref{sec:Chev_corr}. The attractive electron-hole interaction is chosen to be sufficiently large, such that $\psi_\kk(\pbold)$ describes a well-defined, localized bound state. Our approach fully accounts for the composite nature of the exciton and relies exclusively on microscopic interactions without introducing an effective exciton-electron coupling. To study the exciton dynamics, we compute its propagator:
\begin{equation}
    \mathcal{G}_X(\kk, \omega) =  \bra{GS} x_\kk \frac{1}{\omega + i \eta - H} x_\kk^\dagger \ket{GS} \label{eq:Xprop}
\end{equation}
by projecting the Hamiltonian Eq.~\eqref{eq:Hamiltonian} to a subspace that includes states with a single exciton as well as states containing a single exciton accompanied by a particle-hole excitation of the electronic ground state~\cite{chevyUniversalPhaseDiagram2006}. This procedure defines the following basis states: 
\begin{subequations}\label{eq:ChevybasisMain}
\begin{align}
    \ket{n=0} &= \ket{X_\pbold} \equiv x^\dagger_\pbold \ket{GS} \label{eq:ChevybasisMain_a}, \\
    \ket{n>0} &= \ket{C^\pbold_{\kk \alpha \qb \beta}} \equiv x^\dagger_{\pbold+\qb-\kk} c^\dagger_{\alpha \kk} c_{\beta \qb} \ket{GS} \label{eq:ChevybasisMain_b}. 
\end{align}
\end{subequations}
We compute the matrix elements of the Hamiltonian in this basis; see Appendix~\ref{sec:Chevy}.
This approximation~\cite{chevyUniversalPhaseDiagram2006} has proven to be a reliable framework for capturing the many-body physics of an impurity immersed in a bath~\cite{combescotNormalStateHighly2008, amelioPolaronSpectroscopyInteracting2024, amelioPolaronFormationInsulators2024}.

\section{Low and high-density regime}\label{sec:lowhigh}
To connect our results with the established theory on exciton-polaron physics in the continuum~\cite{sidlerFermiPolaronpolaritonsChargetunable2017, imamogluExcitonpolaronsTwodimensionalSemiconductors2020}, we first consider the low and high-filling regimes of the electron band, close to $\nu=0$ and $\nu=2$, where $\nu$ corresponds to the average number of electrons per site. In this regime, the electronic ground state $\ket{GS}$ is a Fermi liquid, for which we treat the repulsive electron-electron interactions on a mean-field level, such that their only effect is to renormalize the electron dispersion $\eps^e_\kk$. Our model correctly captures the formation of an attractive and repulsive polaron upon electron doping, as seen in the zero-momentum exciton spectral function $\mathcal{A}_X(\omega) = -\frac{1}{\pi}\Im \mathcal{G}_X(\kk=0, \omega)$; Fig.~\ref{fig:1}(b). We found that including spin is essential for accurately describing the formation of the attractive polaron in the low-density regime.
This is because the formation of the trion, a bound state of two electrons and a hole, requires the electrons to form a singlet.
By increasing the electron density, the spectral weight shifts from the repulsive to the attractive polaron. 
Our purely electronic model enables the study of how the energies of the two resonances shift with electron filling as a function of the microscopic interactions. Generically, stronger repulsive electron-electron interactions cause a blue shift, while enhanced attractive electron-hole interactions result in a red shift of the resonances. We also explore the impact of different stacking configurations in heterobilayers. For an H-stacked bilayer, the moir\'e sites of the electrons are shifted relative to the hole site. Consequently, the electron component of the exciton wavefunction is predominantly distributed over three sites, all equidistant from the hole site~\cite{wangIntercellMoireExciton2023}. The more delocalized nature of the exciton wave function favors the formation of interlayer attractive polarons, as observed in photoluminescence experiments~\cite{liuSignaturesMoireTrions2021, wangMoireTrionsMoSe22021, wangIntercellMoireExciton2023, upadhyayGiantEnhancementExciton2024}.
For similar attractive interactions, we find that both AP and RP resonances experience a stronger redshift as a function of filling for H-stacking compared to R-stacking, especially in the high-density regime; see Fig.~\ref{fig:1}(b).

We give an order-of-magnitude estimate for the interaction strengths based on WSe$_2$/WS$_2$ in Appendix~\ref{app:estimate}. Due to screening, the actual interaction strength in experiments will depend sensitively on the filling. Such a quantitative computation is beyond the scope of our work. Instead, our qualitative predictions are valid over a wide parameter regime, as exemplified by parameter scans in Appendix~\ref{app:estimate}.

\begin{figure}
\begin{center}
\includegraphics[width=0.99\linewidth]{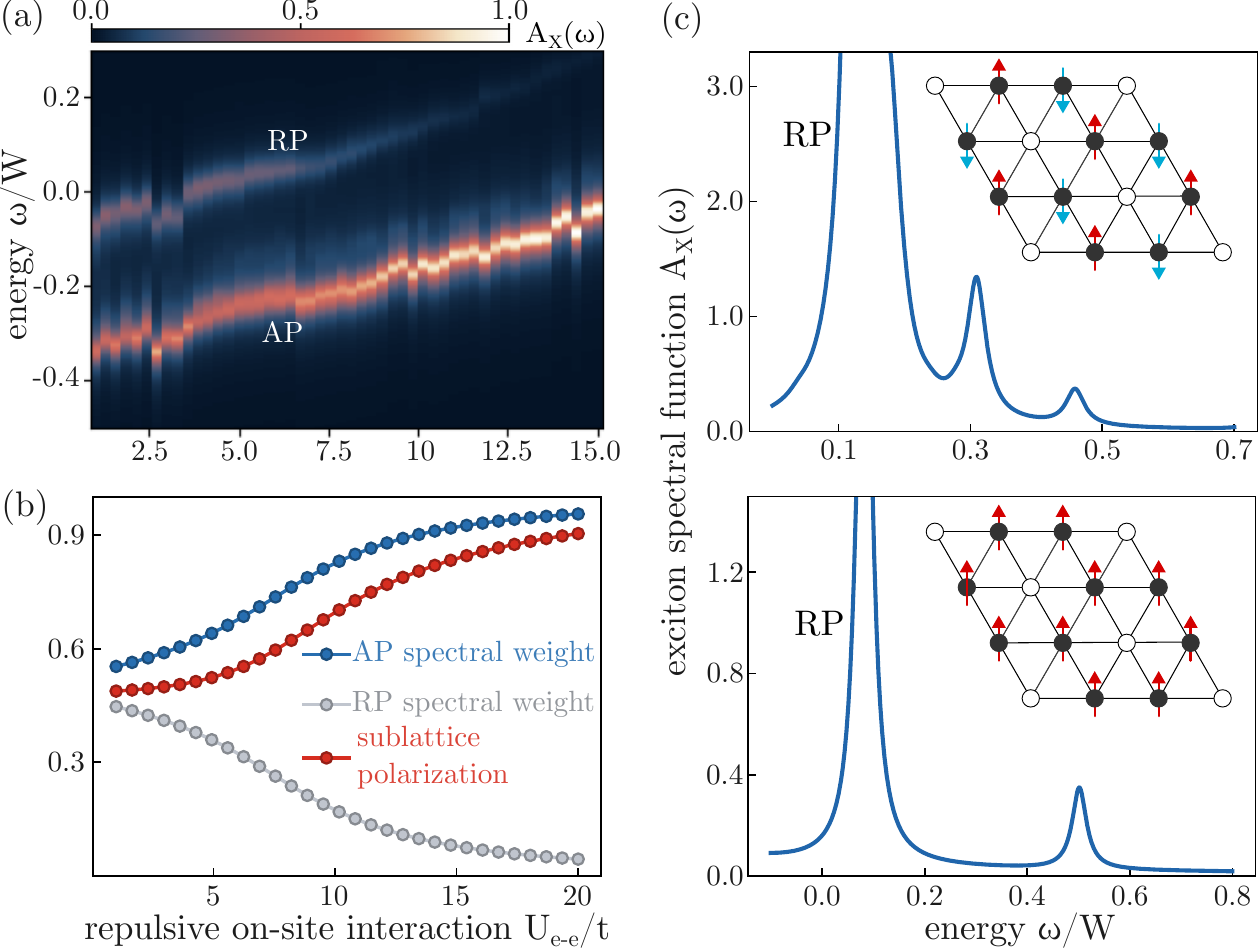}
\caption{\textbf{Polaron formation near correlated insulators.} \textbf{(a)} Zero-momentum exciton spectral function for the $\nu=2/3$ generalized Wigner crystal ($U_{e-h}=-35t, V_{e-e}/U_{e-e} = 1/6$, R-stacking). \textbf{(b)} The spectral weight of the AP increases with sublattice polarization $\Delta n = (n_A  +  n_B  - n_C)/2$, an order parameter for the crystal. Here $ n_{A/B}$ is the average occupation of sublattice $A/B$, forming the crystal, and $n_C$ is the occupation of the third sublattice, which is empty for an ideal crystal. \textbf{(c)} We observe two Umklapp peaks for the antiferromagnet, while there is only one Umklapp peak for the ferromagnet ($U_{e-e} = 12t$, the ferromagnet is stabilized by a direct exchange $X_{e-e} = 0.8t$).
}
\label{fig:2}
\end{center}
\end{figure}

\section{Correlated Insulators} \label{sec:corrins}
So far, we have assumed the charge carriers to be in a metallic state. However, in intermediate density regimes, correlated insulators are observed at commensurate fillings~\cite{regan_mott_2020, shimazakiStronglyCorrelatedElectrons2020, wang_correlated_2020, xu_correlated_2020, huang_correlated_2021, li_imaging_2021,  tang_evidence_2023}. We first focus on the generalized Wigner crystal at $\nu = 2/3$. The exciton-polaron spectral function of spinless generalized Wigner crystals was discussed in Refs.~\cite{amelioPolaronSpectroscopyInteracting2024, amelioPolaronFormationInsulators2024} using conventional Bose-Fermi models. We treat the repulsive electron-electron interactions on a mean-field level, allowing for the spontaneous breaking of the discrete translation symmetry. 
Concretely, we increase the unit cell to include three sites, allowing for the expected spin- and charge order at $\nu = 2/3$. After mean-field decoupling of the electron-electron interaction, the electron Hamiltonian Eq.~\eqref{eq:Ham_elec} is
\begin{equation}
    H_e = \displaystyle\sum_{\kk}\displaystyle\sum_{\lambda=1}^6 \epsilon^\lambda_\kk \gamma^\dagger_{\lambda \kk} \gamma_{\lambda \kk}.
\end{equation}
The Hartree-Fock quasiparticles $\gamma^\dagger_{\lambda \kk}$ with dispersion $\epsilon^\lambda_\kk$ are obtained by self-consistently solving the electronic mean-field problem; see Appendix~\ref{sec:Chev_corr}. To define the momenta of the top and bottom layers with respect to the same translations, we also enlarge the unit cell for the holes.
After expressing Eqs.~\eqref{eq:ChevybasisMain} in this new basis, we determine the exciton wavefunction by solving the two-particle Schr{\"o}dinger equation in the presence of the correlated-insulator ground state $\ket{GS}$. The resulting wavefunction describes the long-lived, lowest-energy exciton state to which optically created excitons relax. However, our approach also allows us to target exciton bound states at higher energies.

For the generalized Wigner crystal at $\nu=2/3$, both the RP and AP branch blueshift with increasing electron-electron interaction $U$, which scales with the charge gap; Fig.~\ref{fig:2}(a). As the charge gap increases and the crystal order becomes more established, the spectral weight of the AP increases; see Fig.~\ref{fig:2}(b). Hence, the AP spectral weight peaks when the electrons are fully localized.

Optical measurements provide the exciton spectral function at zero momentum, as momentum transfer by optical light is negligible. States with a periodic charge distribution coupled to excitons, however, give rise to additional optically-active Umklapp resonances~\cite{shimazakiOpticalSignaturesCharge2021, Smolenski_ObservationWigner_2021}, at which a finite momentum state at a reciprocal lattice vector is backfolded to zero quasi-momentum. 
Due to the valley-Zeeman effect, the Umklapp peaks were also predicted to exhibit distinct signatures for antiferromagnetic and ferromagnetic spin order, offering a potential method to differentiate between the two~\cite{Salvador_OpticalSignatures_2022}. The spin order of experimentally observed generalized Wigner crystals is still unclear, as definitive experimental probes are lacking~\cite{pichlerProbingMagnetismMoire2024}. Both antiferromagnetic and ferromagnetic couplings are possible, depending on the strength of direct exchange interactions, which favor ferromagnetism~\cite{Morales-Duran_NonlocalInteractions_2022}. 

For the generalized Wigner crystal at $\nu=2/3$, which forms a honeycomb charge order, we observe Umklapp peaks near the RP resonance. The number of Umklapp peaks depends on the symmetry of the state~\cite{Salvador_OpticalSignatures_2022}. A generalized Wigner crystal with ferromagnetic spin order has $C_6$ symmetry, allowing only a single Umklapp peak. On the other hand, an antiferromagnet has a lower $C_3$ symmetry and more Umklapp peaks are allowed by symmetry. These symmetry arguments perfectly agree with our results, which consist of only a single peak for ferromagnetic spin order and further splitting of the peaks for antiferromagnetic spin order; see Fig.~\ref{fig:2}(c). We tune from an antiferromagnetic to a ferromagnetic crystal by introducing a finite nearest-neighbor direct exchange interaction; see Appendix~\ref{sec:Chev_corr} and Ref.~\cite{Morales-Duran_NonlocalInteractions_2022}.
Our model also predicts Umklapp peaks for the AP, however, with much weaker spectral weight than for the RP.

\begin{figure}
\begin{center}
\includegraphics[width=1\linewidth]{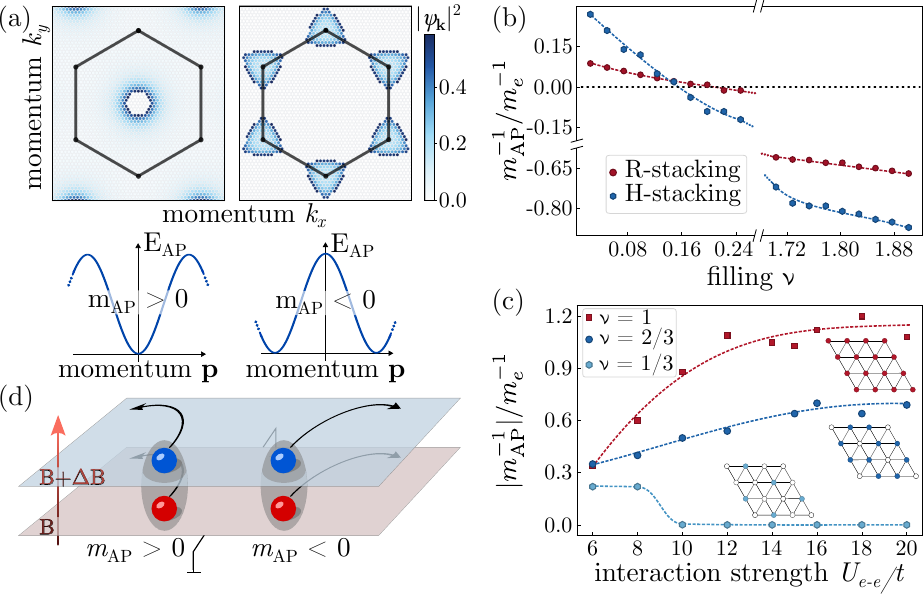}
\caption{\textbf{Mass renormalization of attractive polaron.} 
\textbf{(a)} Exciton wavefunction $\psi_\kk(\pbold=0)$ at low (left) and high (right) densities ($\nu=0.06$ and $\nu=1.67$, respectively). In the high-density regime, the exciton wavefunction is localized near the band maxima at the Brillouin zone corners, leading to a negative exciton-polaron mass. \textbf{(b)} Inverse of the AP mass $m^{-1}_{AP}$ relative to the electron mass $m^{-1}_e = 3t$, in the low and high filling regime, with a Fermi sea as the ground state. \textbf{(c)} Inverse mass as a function of the interaction strength $U$ for correlated insulators at $\nu=1$, $\nu=2/3$, and $\nu=1/3$ in R-stacking. \textbf{(d)} The sign change of the polaron mass can be detected by a Hall-type measurement, where a magnetic field with an out-of-plane gradient is applied. The resulting Hall current is sensitive to the sign of the polaron mass. Details of the parameters can be found in Appendix~\ref{app:determining}.}
\label{fig:3}
\end{center}
\end{figure}

\section{Large mass renormalization}
Immersing an exciton in a correlated electronic bath can lead to a strong renormalization of its properties. 
We now theoretically investigate the mass renormalization of the AP
as a function of the electron filling from the momentum-dependent exciton spectral function. 
We define the mass $m$ of a particle with dispersion $\varepsilon_\kk$ as $m^{-1} = \partial^2 \varepsilon_\kk/ \partial \kk^2 |_{\kk=0}$. For a particle hopping on a triangular lattice, the mass is given by $m=3/W$, where $W$ is the bandwidth of $\varepsilon_\kk$. We use this relation to determine the mass of the AP by calculating its bandwidth. We find a very strong mass renormalization of the polaron, consistent with recent diffusion experiments~\cite{upadhyayGiantEnhancementExciton2024, yanAnomalouslyEnhancedDiffusivity2024}. We emphasize that this mass renormalization for interlayer-excitons in moir\'e lattices is orders of magnitude larger than the typical mass renormalization for polarons in the continuum~\cite{combescotNormalStateHighly2007, nascimbeneCollectiveOscillationsImbalanced2009, navonEquationStateLowTemperature2010}. 
In a system with continuous Galilean invariance, the effective mass of a bound state is independent of its binding energy. In contrast, in a lattice, where center-of-mass motion does not decouple, the effective mass depends on the binding energy; and second-order perturbation theory predicts that the exciton mass is proportional to the binding energy.

At low electron densities, increasing the filling provides more electronic excitations for the exciton to bind with, reducing its mobility; Fig.~\ref{fig:3}(b). In the high-density regime where the electron band is nearly full, available excitations decrease again, leading to weaker mass renormalization. Additionally, at sufficiently high densities, new hopping channels emerge, allowing the hole to hop independently and rebind with a distant electron, enhancing mobility, referred to as non-monogamous hopping in Ref.~\cite{upadhyayGiantEnhancementExciton2024}. This trend resembles exciton mobility measured in recent experiments~\cite{upadhyayGiantEnhancementExciton2024, yanAnomalouslyEnhancedDiffusivity2024}.

The structure of the exciton wavefunction $\psi_\kk(\pbold)$ also greatly affects the exciton-polaron mobility, leading to a higher mobility in H-stacking than in R-stacking due to the less-localized exciton wavefunction in H-stacking; see Fig.~\ref{fig:3}(b). Relatively, the difference in mobility between R- and H-stacking is more substantial in the low-density regime. Another striking effect stemming from the non-trivial structure of the exciton wavefunction is a sign change in the effective polaron mass at a critical filling $\nu_0$. At low densities, the Fermi surface is close to the band minimum with positive curvature, resulting in a positive exciton mass. At high densities, however, available states for exciton formation lie near the band maxima at the Brillouin zone corners; see Fig.~\ref{fig:3}(a). There, the negative band curvature leads to a negative exciton mass. The critical density $\nu_0$ at which this curvature effect becomes significant is quite low, with the precise value depending on the microscopic interactions. 
We propose to detect this sign change in the effective polaron mass in a Hall-type experiment. Since the charge carriers in the bottom layer are bound within excitons, the current in the bottom layer is directly proportional to the exciton-polaron transport. Applying a magnetic field with a spatial out-of-plane gradient induces a finite Lorentz force on the exciton, generating a measurable Hall current. Crucially, the Hall current is sensitive to the sign of the polaron mass, see e.g.~\cite{Ziman_1972}, providing an experimental signature of the sign change; see Fig.~\ref{fig:3}(d). 
While a magnetic field gradient in the out-of-plane direction is necessary for the bare exciton to experience a finite Lorentz force, the AP is a charged quasiparticle and thus couples directly to the magnetic field and no gradient is needed.

We now study the mass renormalization in the vicinity of correlated insulators. Both for $\nu = 2/3$ and $\nu=1$, we observe a very small effective mass, corresponding to a very large mobility; see Fig.~\ref{fig:3}(c). 
Assuming that the attractive electron-hole interaction dominates over the repulsive electron-electron interaction $|U_{e-h}| > U_{e-e}$, the exciton forms with an electron on top of the crystal, resulting in a weakly bound state. This results in a high mobility for $\nu=2/3$ and $\nu=1$, as the exciton hopping scales with the inverse of the binding energy. Since the exciton can only hop on top of the crystal, the mobility is smaller in the $\nu=2/3$ than the $\nu=1$ state, as there are fewer available hopping sites.
Polaron mobility increases with repulsive electron-electron interaction until charge and spin order are well established; at this point, the mobility starts to saturate. In contrast, for $\nu=1/3$ generalized Wigner crystals, the exciton is effectively localized as there are no nearest-neighbor sites for it to hop to while remaining on the crystal. 

Depending on the energy scales, for the generalized Wigner crystals, it is possible to obtain excitons that occupy an empty site. Such a configuration is favored for weaker on-site $|U_{e-h}| \lesssim U_{e-e}$ and stronger long-range $V_{e-h}$ attractive interactions. This enhances mobility for the $\nu=1/3$ state but drastically reduces it for the $\nu=2/3$ state, demonstrating how different microscopic interactions can dramatically alter the exciton mobility near correlated insulators. Further details are provided in Appendix~\ref{app:determining}.

We stress that having a purely electronic model is essential for capturing the strong mass renormalization for excitons in moir\'e lattices. Treating excitons as tightly-bound bosons fails to account for the diverse processes arising from their non-trivial wavefunction. By contrast, our model, Eq.~\eqref{eq:Hamiltonian}, inherently incorporates these effects.

\section{Conclusion and Outlook}
We have developed an effective, purely electronic model to describe exciton-polaron formation in moir\'e heterostructures. Our model 
highlights key mechanisms of stacking dependence and mass renormalization of the exciton polarons, consistent with recent experimental observations of giant exciton mobilities~\cite{upadhyayGiantEnhancementExciton2024, yanAnomalouslyEnhancedDiffusivity2024}. While our work was inspired by and is directly relevant to experiments in WSe$_2$/WS$_2$ heterobilayers, our model is not tailored to any specific material. Beyond TMDs, the model may also be relevant for exciton formation in other strongly correlated Mott insulators, such as $\text{Sr}_2\text{IrO}_4$~\cite{mehioHubbardExcitonFluid2023}. An exciting direction for future research is to extend our approach to explore polaron formation in slightly doped Mott insulators, where the exciton is dressed by collective spin modes. Furthermore, our framework could be applied to other exotic states, such as fractional quantum anomalous Hall states~\cite{cai_SignaturesFractional_2023, zeng_ThermodynamicEvidence_2023, park_ObservationFractionally_2023, xu_ObservationInteger_2023} or anomalous Hall crystals~\cite{dong_AnomalousHall_2023, dongTheoryQuantumAnomalous2024}, using a parton mean-field state as the electronic ground state.

\section{Acknowledgements}
We thank Wilhelm Kadow, Tsung-Sheng Huang, Pranshoo Upadhyay, Clemens Kuhlenkamp, and Ivan Amelio for insightful discussions. We acknowledge support from the Deutsche Forschungsgemeinschaft (DFG, German Research Foundation) under Germany’s Excellence Strategy–EXC–2111–390814868, TRR 360 – 492547816 and DFG grants No. KN1254/1-2, KN1254/2-1, the European Research Council (ERC) under the European Union’s Horizon
2020 research and innovation programme (grant agreement No 851161), the European Union (grant agreement No 101169765), as well as the Munich Quantum Valley, which is supported by the Bavarian state government with funds from the Hightech Agenda Bayern Plus. This research was supported in part by grant NSF PHY-2309135 to the Kavli Institute for Theoretical Physics (KITP).

\section{Data availability}
Data and codes are available upon reasonable request on Zenodo~\cite{zenodo}.
}

\appendix
\section{Exciton wavefunction}\label{sec:ExcWave}

\begin{figure*}
\begin{center}
\includegraphics[width=0.99\linewidth]{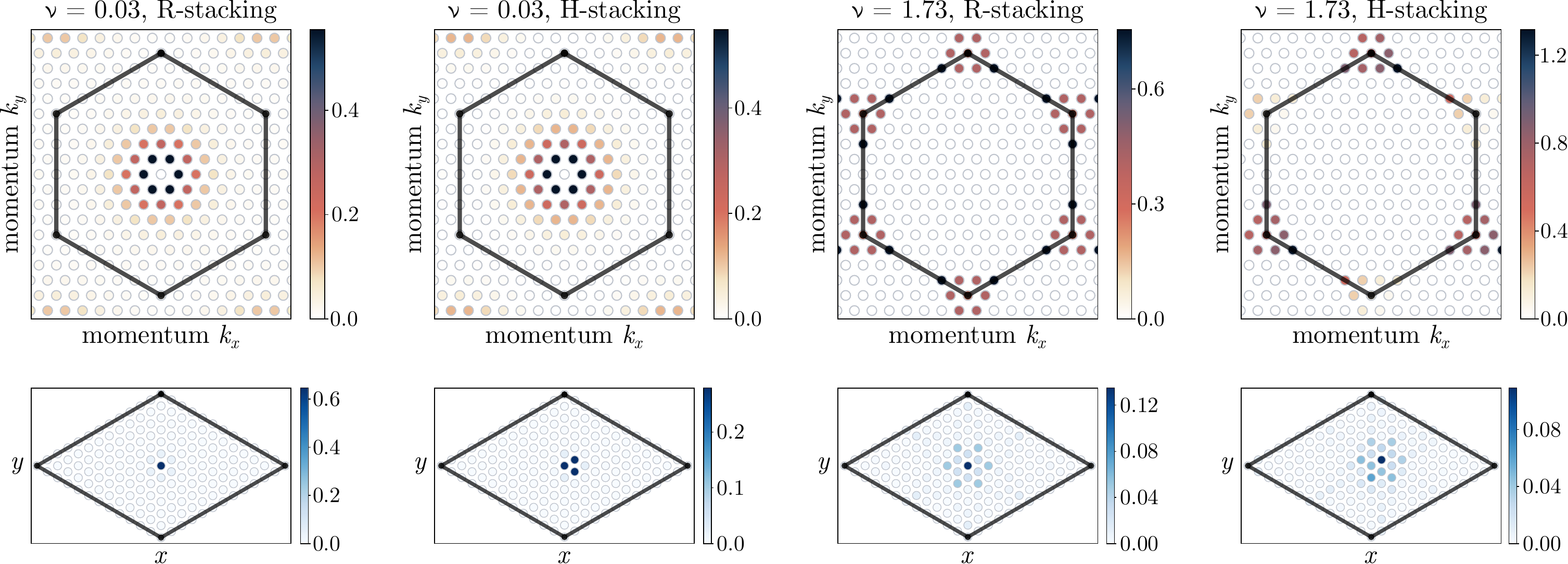}
\caption{\textbf{Exciton wavefunction.} Top: Zero total momentum exciton wavefunction $\psi_\kk(\pbold=0)$ for different stackings and electron densities $\nu$. The boundaries of the Brillouin zone are highlighted. Bottom: Fourier transform of $\psi_\kk(\pbold=0)$, showing localization in real space. The exciton wavefunction is most strongly localized for R-stacking at low densities. $U_{e-h}=-15t$ with all other interactions set to zero. System size $N=12\times12$.
}
\label{fig:Supp_exWF}
\end{center}
\end{figure*}

Before tackling the many-body problem of inserting a single exciton into some electronic ground state $\ket{GS}$, we solve the two-body problem of an electron-hole pair. We define the exciton wavefunction with total momentum $\pbold$ as
\begin{equation}
    \ket{X_\pbold} = x^\dagger_\pbold \ket{GS} = \displaystyle\sum_\kk \psi_{\kk}(\pbold) c^\dagger_{\uparrow \pbold - \kk} h^\dagger_{\uparrow \kk} \ket{GS}. \label{eq:exitonWavefuncApp}
\end{equation}
where the ground state 
\begin{equation}
    \ket{GS} = \displaystyle\prod_{|\kk|< k_F, \sigma} c^\dagger_{\sigma \kk} \ket{0}
\end{equation}
is assumed to be a Fermi sea with the Fermi momentum $k_F$ fixing the filling $\nu$. Here $\ket{0}$ is the vacuum state, corresponding to charge neutrality. Note that we chose to fix the spin of the electron and hole forming the exciton when working with a Fermi sea as the electronic ground state. We derive an equation for $\psi_\kk(\pbold)$ in Eq.~\eqref{eq:exitonWavefuncApp}, by projecting the Schr{\"o}dinger equation $H\ket{X_\pbold} = E^X_\pbold \ket{X_\pbold}$ onto $\bra{0} h_\kk c_{\pbold-\kk}$, leading to:
\begin{equation}
\begin{aligned}
    (\eps_{\pbold-\kk}^e &+ \eps_{\kk}^h +\nu V_{\qb=0}) \nbar_{\uparrow \pbold-\kk} \psi_{\kk}(\pbold)
       \\ +&\displaystyle\sum_{\kk'} V_{\kk'-\kk} \nbar_{\uparrow \pbold-\kk} \nbar_{\uparrow \pbold-\kk'} \psi_{\kk'}(\pbold) = E^X_\pbold \psi_\kk(\pbold)  \label{eq:excitonSchro}
\end{aligned}
\end{equation}
with $\nbar_{\sigma \kk} = 1-n_{\sigma \kk} = 1-\langle{c_{\sigma \kk}^\dagger c_{\sigma \kk}\rangle}$. We treat the repulsive electron-electron interaction purely on a mean-field level, renormalizing the electron dispersion $\eps_\kk^e$. The free electron and hole dispersions are taken as
\begin{equation}
    \eps_{\kk}^{e/h} = -2t [\cos(\kk\cdot \mathbf{a}_1) + \cos(\kk\cdot \mathbf{a}_2) + \cos(\kk\cdot \mathbf{a}_3)]
\end{equation}
with lattice translation vectors $\mathbf{a}_{1/2} =  (\pm \sqrt{3}/2, 1/2)^T$ and $\mathbf{a}_3 = (0, -1)^T$, corresponding to a particle hopping on a triangular lattice with hopping parameter $t$.
We use the lowest-energy solution of Eq.~\eqref{eq:excitonSchro} to define the exciton operator $x_\pbold^\dagger$ though Eq.~\eqref{eq:exitonWavefuncApp}. For R-stacking, the exciton wavefunction is strongly localized in real space, while for H-stacking, it is spread over three sites; see Fig.~\ref{fig:Supp_exWF}. At very high electron densities $\nu$, most states are already occupied, such that $\psi_\kk(\pbold)$ is only non-zero for $\kk$ close to the corners of the Brillouin zone, leading to a less localized exciton wavefunction in real space.

\section{Attractive electron-hole interactions and stacking dependence} 

\label{sec:app:attracInt}
The top layer hosting electrons and the bottom layer hosting holes are coupled through attractive electron-hole interactions Eq.~\eqref{eq:Ham_eh}, responsible for forming the exciton as an electron-hole bound state. We are working with on-site and nearest-neighbor density-density interactions:
\begin{equation}
    H_{e-h} =  U \displaystyle\sum_{i, \sigma, \sigma'} c^\dagger_{\sigma i} h^\dagger_{\sigma' i} h_{\sigma' i} c_{\sigma i} + V\displaystyle\sum_{\substack{\langle i, j \rangle \\  \sigma, \sigma'}} c^\dagger_{\sigma i} h^\dagger_{\sigma' j} h_{\sigma' j} c_{\sigma i},
\end{equation}
with $U, V < 0$. Note that for H-stacking, the electron and hole sites are shifted with respect to each other, such that for each hole site, there are three equally distant electron sites, which we consider to be ``local''. After taking a Fourier transform, we obtain
\begin{equation}
        H_{e-h} = \displaystyle\sum_{\kk \kk' \qb}\displaystyle\sum_{\sigma \sigma'} V_\qb c^\dagger_{\sigma, \kk + \qb} h^\dagger_{\sigma', \kk'-\qb} h_{\sigma', \kk'}c_{\sigma \kk}, \label{eq:Ham_eh_app}
\end{equation}
where the shape of $V_\qb$ depends on the stacking. For R-stacking, one has
\begin{equation}
    V_\qb =  \frac{U}{N} + \frac{V}{N} ( e^{-i \mathbf{a}_1 \cdot \qb} + e^{-i \mathbf{a}_2 \cdot \qb} + e^{-i \mathbf{a}_3 \cdot \qb}),
\end{equation}
with lattice translation vectors $\mathbf{a}_{1/2} =  (\pm \sqrt{3}/2, 1/2)^T$ and $\mathbf{a}_3 = (0, -1)^T$ and $N$ is the number of unit cells. On the other hand, for H-stacking, we find
\begin{equation}
    \begin{aligned}
    V_\qb = &\frac{U}{N}(1 + e^{-i \mathbf{a}_2 \cdot \qb} + e^{+i \mathbf{a}_1 \cdot \qb}) \\+& \frac{V}{N} (e^{+i( \mathbf{a}_1-\mathbf{a}_2) \cdot \qb} + e^{-i \mathbf{a}_3 \cdot \qb} + e^{+i \mathbf{a}_3 \cdot \qb}).
    \end{aligned}
\end{equation}

\section{Estimate for the interaction strength}
\label{app:estimate}

\begin{figure}
\begin{center}
\includegraphics[width=0.99\linewidth]{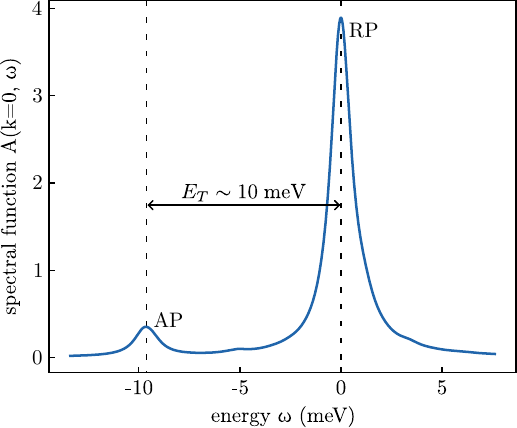}
\caption{\textbf{Trion binding energy}. We fix the interaction strength in the main text by calculating the binding energy of the trion, which corresponds to the energy difference between the AP and RP in the low-density regime $\nu\rightarrow 0$. Here we use H-stacking, system size $N=12\times 12$ and $U_{e-h}=15t$ and density $\nu = 1/36$, resulting in a trion energy of $E_T \sim 10\; \text{meV}$, consistent with experiments~\cite{upadhyayGiantEnhancementExciton2024}. The hopping parameter is estimated to be $t\sim 1$ meV. All other interaction strengths are set to zero.
}

\label{fig:trionEnergy}
\end{center}
\end{figure}

Our model and the predictions we derive from it are primarily qualitative, with the main goal being to provide a simple model capturing the essential mechanism underlying exciton-polaron formation in moir\'e heterostructures. Nevertheless, we connect to experiments by giving a rough estimate for the interaction parameters used in the main text. We assume a WSe$_2$/WS$_2$ heterostructure, as used in Ref.~\cite{upadhyayGiantEnhancementExciton2024}. At zero twist angle, the moir\'e lattice constant is given by $a_M = a/\delta \sim 7\;\text{nm}$, where we take $a=0.3297\;\text{nm}$ as the atomic lattice constant for WSe$_2$ and $\delta$ is its lattice mismatch with WS$_2$. Taking an effective electron mass $m^* = 0.5 m_e$, we obtain a rough estimate for the hopping parameter of the electrons on a triangular lattice
\begin{equation}
    t \sim \frac{\hbar^2}{3 a_M^2 m^* } \sim 1\;\text{meV}.
\end{equation}
We want to estimate the effective interaction strengths in the low-density regime, where we assume to have a Fermi liquid as the electronic ground state. In that case, the only effect of electron-electron interactions is to renormalize the bare parameters. We capture this by treating the electron-electron interactions on a mean-field level. Within this approximation, on-site interactions just introduce a density-dependent on-site potential
\begin{equation}
    H_U = U_{e-e} \displaystyle\sum_{i} c^\dagger_{\uparrow i} c^\dagger_{\downarrow i} c_{\downarrow i} c_{\uparrow i} \approx U_{e-e} n_e \displaystyle\sum_{i,\sigma} c_{\sigma i}^\dagger c_{\sigma i} \label{eq:U_MeanField}
\end{equation}
with $n_e=\langle c_{\sigma i}^\dagger c_{\sigma i} \rangle$ the average density. Similarly, nearest-neighbor density-density interactions result in an additional shift of the on-site potential by $4z n_e V_{e-e}$, where $z=6$ is the number of nearest-neighbors on the triangular lattice. We note that we always work with a fixed number of particles. Consequently, a shift in the on-site potential should be understood as a shift of the electron band relative to the hole band, influencing the exciton energy. Additionally, a finite nearest-neighbor interaction renormalizes the electron hopping $t \rightarrow t+\bar{t}$, with
\begin{equation}
    \bar{t} = \frac{V_{e-e}}{N} \displaystyle\sum_{\kk} e^{i \kk \cdot \mathbf{a}} \langle c_{\sigma \kk}^\dagger c_{\sigma \kk}\rangle
\end{equation}
where $\mathbf{a}$ is a lattice vector, connecting nearest-neighbor sites of the triangular lattice. For low densities, we now take the repulsive interactions to be zero, which can be equivalently understood as working with the Fermi-liquid quasiparticles, instead of the bare electrons. Hence, also the value of the attractive interactions $U_{e-h}$, which we now estimate from experiments, should be thought of as the effective renormalized interaction strength. We fix $U_{e-h}$ through the trion binding energy, which was measured in Ref.~\cite{upadhyayGiantEnhancementExciton2024} as a redshift of the order of $E_T \sim10\;\text{meV}$ in the photoluminescence spectrum, when tuning away from charge neutrality $\nu=0$. In our calculation, this redshift corresponds to the difference in energy between the AP and RP resonances at low densities; see Fig.~\ref{fig:trionEnergy}. Assuming, for simplicity, vanishing nearest-neighbor attraction, we find $U_{e-h} = 15t$ is consistent with $E_T \sim 10$ meV.

\begin{figure*}
\begin{center}
\includegraphics[width=0.89\linewidth]{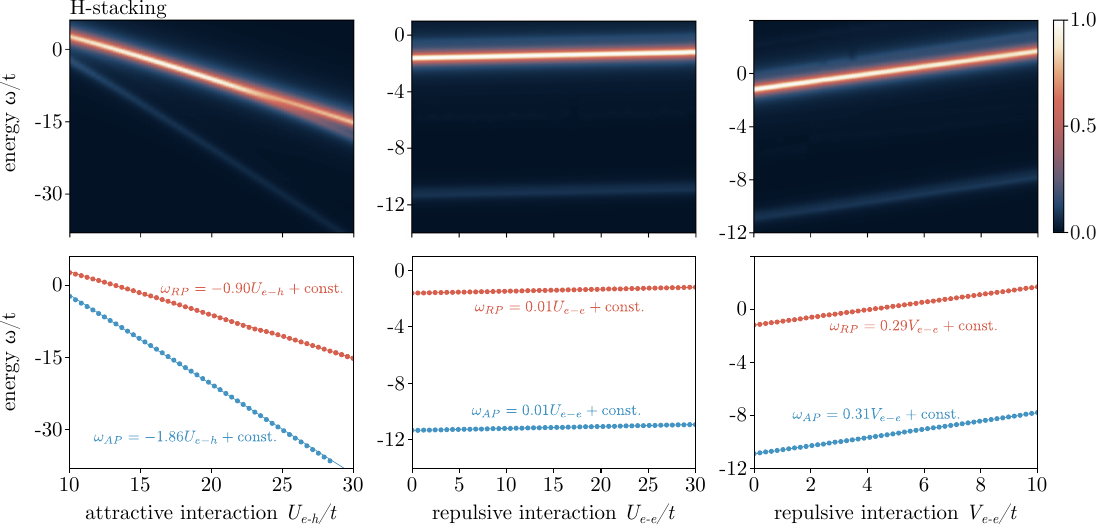}
\caption{\textbf{Parameter scans for H-stacking}. Top panels show the zero-momentum spectral function as a function of various interaction strengths. Bottom panels show the extracted RP and AP resonances and the linear fit to them. Left: Tuning the attractive electron-hole interaction $U_{e-h}$, with all other interactions set to zero. The AP redshifts approximately twice as fast as the RP, consistent with the interpretation of the AP as a charged exciton. Center: Tuning the repulsive on-site interaction $U_{e-e}$, with $U_{e-h}=15t$ fixed and all other interactions zero. On a mean-field level, the only effect of $U_{e-e}$ is to provide a constant energy shift. Right: Tuning the repuslive nearest-neighbor interaction $V_{e-e}$, with $U_{e-h}=15t$ and $U_{e-e}=30t$ fixed. $N=12\times 12$, $\nu=1/72 \approx 0.014$ 
}
\label{fig:scanH}
\end{center}
\end{figure*}

\begin{figure*}
\begin{center}
\includegraphics[width=0.89\linewidth]{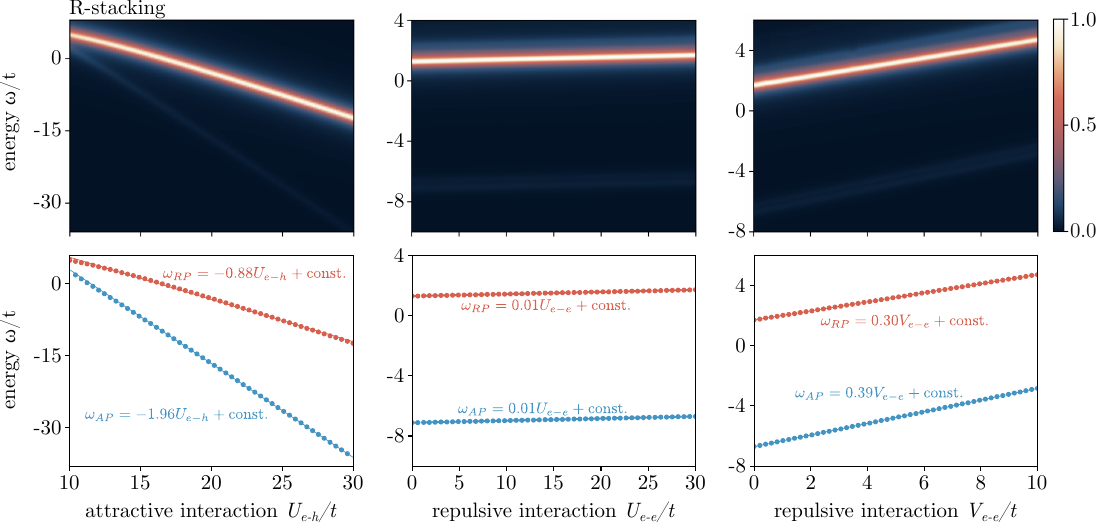}
\caption{\textbf{Parameter scans for R-stacking}. Top panels show the zero-momentum spectral function as a function of various interaction strengths. Bottom panels show the extracted RP and AP resonances and the linear fit to them. Left: Tuning the attractive electron-hole interaction $U_{e-h}$, with all other interactions set to zero. The AP redshifts approximately twice as fast as the RP, consistent with the interpretation of the AP as a charged exciton. Center: Tuning the repulsive on-site interaction $U_{e-e}$, with $U_{e-h}=15t$ fixed and all other interactions zero. On a mean-field level, the only effect of $U_{e-e}$ is to provide a constant energy shift. Right: Tuning the repuslive nearest-neighbor interaction $V_{e-e}$, with $U_{e-h}=15t$ and $U_{e-e}=30t$ fixed. $N=12\times 12$, $\nu=1/72 \approx 0.014$ 
}
\label{fig:scanR}
\end{center}
\end{figure*}

We use this value of $U_{e-h}$ in Fig.~\ref{fig:1}. Note that the actual interaction strength will be strongly renormalized as a function of density due to screening. Accurately capturing this is beyond the scope of this work. Instead, we assume a constant interaction strength in the low and high-density regimes and show parameter scans over different interaction strengths to argue that our results are qualitatively stable; see Figs.~\ref{fig:scanH},\ref{fig:scanR}. In the Fermi-liquid regime, increasing repulsive interactions $U_{e-e}$ and $V_{e-e}$ lead to a linear blueshift, as a consequence of the density-dependent on-site potential Eq.~\eqref{eq:U_MeanField}. At low densities, the effect of increasing $U_{e-e}$ and $V_{e-e}$ is marginal, which is why we have neglected it in the discussion of Fig.~\ref{fig:1} in the main text.

Increasing the attractive electron-hole interaction $U_{e-h}$ induces a redshift of both the RP and AP resonances. We find that the rate at which the AP redshifts is approximately twice as fast as the redshift of the RP. This is consistent with the interpretation of the AP as a charged exciton. Neglecting kinetic energy we expect a naive binding energy of $\omega_{AP} \sim 2 U_{e-h}$, while the energy of the RP, which is continuously connected to the bare exciton, should scale as $\omega_{RP} \sim U_{e-h}$. This is consistent with what we find; Figs.~\ref{fig:scanH},\ref{fig:scanR}.

\section{Correlated insulators} \label{sec:Chev_corr}
We enlarge the unit cell to include three sublattices to describe the correlated insulators at $\nu=1/3$, $\nu=2/3$, and $\nu=1$. The lattice translation vectors with respect to the new unit cell are $\mathbf{b}_1 =  \mathbf{a}_1 - \mathbf{a}_3$ and $\mathbf{b}_2 = \mathbf{a}_2 - \mathbf{a}_3$. Furthermore, we define $\mathbf{b}_0 = \mathbf{a}_1 - \mathbf{a}_2$, with the translation vectors $\mathbf{a}_i$ of the original lattice. We treat the repulsive electron-electron interactions on a mean-field level, defining $\mfp_{\sigma \sigma'}^{i j} := \bexpval{c_{\sigma i}^\dagger c_{\sigma' j}}$. 
\begin{widetext}
The electron-electron interaction Hamiltonian reads
\begin{equation}
    H_{e-e} =  U_{e-e} \displaystyle\sum_{i, \sigma, \sigma'} c^\dagger_{\sigma i} c^\dagger_{\sigma' i} c_{\sigma' i} c_{\sigma i} + V_{e-e}\displaystyle\sum_{\substack{\langle i, j \rangle \\  \sigma, \sigma'}} c^\dagger_{\sigma i} c^\dagger_{\sigma' j} c_{\sigma' j} c_{\sigma i} + X_{e-e} \displaystyle\sum_{\substack{\langle i, j \rangle \\  \sigma, \sigma'}} c^\dagger_{\sigma i} c^\dagger_{\sigma' j} c_{\sigma' i} c_{\sigma j}.
\end{equation}
After mean-field decoupling, it becomes:
\begin{subequations}
\begin{align}
    H_U &= U_{e-e} \displaystyle\sum_{i, \sigma} (\mfp_{\sigma \sigma}^{i i} c_{\bar{\sigma} i}^\dagger c_{\bar{\sigma} i} - \mfp_{\sigma \bar{\sigma}}^{i i} c_{\bar{\sigma} i}^\dagger c_{\sigma i})
    - U_{e-e} \displaystyle\sum_{i}(\mfp_{\uparrow \uparrow}^{i i} \mfp_{\downarrow \downarrow}^{i i} - \mfp_{\uparrow \downarrow}^{i i} \mfp_{\downarrow \uparrow}^{i i}) \\
    H_V &= V_{e-e} \displaystyle\sum_{\expval{ij}}\displaystyle\sum_{\sigma \sigma'} \big( \mfp_{\sigma \sigma}^{i i} c_{\sigma' j}^\dagger c_{\sigma' j} + \mfp_{\sigma' \sigma'}^{j j} c_{\sigma i}^\dagger c_{\sigma i} - \mfp_{\sigma' \sigma}^{j i} c_{\sigma i}^\dagger c_{\sigma' j} - \mfp_{\sigma  \sigma'}^{i j} c_{\sigma' j}^\dagger c_{\sigma i} - \mfp_{\sigma \sigma}^{i i} \mfp_{\sigma' \sigma'}^{j j} + \mfp_{\sigma' \sigma}^{j i} \mfp_{\sigma \sigma'}^{i j}\big) \numberthis\\
    H_X &= X_{e-e}\displaystyle\sum_{\expval{ij}} \displaystyle\sum_{\sigma \sigma'} \big( \mfp_{\sigma \sigma'}^{i i} c_{\sigma' j}^\dagger c_{\sigma j} + \mfp_{\sigma' \sigma}^{j j} c_{\sigma i}^\dagger c_{\sigma' i} - \mfp_{\sigma \sigma}^{i j} c_{\sigma' j}^\dagger c_{\sigma' i} - \mfp_{\sigma' \sigma'}^{j i} c_{\sigma i}^\dagger c_{\sigma j} - \mfp_{\sigma' \sigma}^{j j} \mfp_{\sigma \sigma'}^{i i} + \mfp_{\sigma \sigma}^{i j} \mfp_{\sigma' \sigma'}^{j i}\big). \numberthis
\end{align}
\end{subequations}
where the nearest-neighbor direct exchange $H_X$ favors ferromagntism~\cite{Morales-Duran_NonlocalInteractions_2022}. The notation $\bar{\sigma}$ implies opposite spin relative to $\sigma$, i.e., $\bar{\uparrow} = \downarrow$ and $\bar{\downarrow} = \uparrow$. In the enlarged unit cell, we bring the mean-field Hamiltonian into bilinear form $H_e = \sum_\kk \Psi_\kk^\dagger H_\kk \Psi_\kk$, with the sublattice spinor $\Psi_\kk = (c_{A \uparrow \kk}, c_{A \downarrow \kk}, c_{B \uparrow \kk}, c_{B \downarrow \kk}, c_{C \uparrow \kk}, c_{C\downarrow \kk})^T$. The matrix $H_\kk$ depends on the mean-field parameters $\mfp_{\sigma \sigma'}^{ab}$, which we determine self-consistently. To define the momenta of the top and bottom layers with respect to the same translations, we also need to enlarge the unit cell for the holes, but since we neglect hole-hole interactions, this only leads to a folding of the hole bands. Concretely, we write the electron and hole operator in the enlarged unit cell in terms of the new eigenbases
\begin{equation}
   c_{\sigma a \kk} = \displaystyle\sum_{\lambda=1}^6 \mathcal{U}^\lambda_{\sigma a}(\kk) \gamma_{\lambda \kk} \qq{and} h_{\sigma a \kk} = \displaystyle\sum_{\lambda=1}^6 \mathcal{W}^\lambda_{\sigma a}(\kk) h_{\lambda \kk}, \label{eq:newbasis}
\end{equation}
where $a\in \{A, B,C\}$ is the sublattice index. Expressed in these new bases, the electron-hole interaction becomes
\begin{equation}
    H_{e-h}= \displaystyle\sum_{\kk \kk' \qb}\displaystyle\sum_{\substack{\mu \mu' \\ \nu \nu'}}V^{\nu' \nu}_{\mu' \mu}(\kk, \kk', \qb) \gamma^\dagger_{\mu', \kk+\qb} h^\dagger_{\nu', \kk'-\qb} h_{\nu, \kk'} \gamma_{\mu, \kk},
\end{equation}
with 
\begin{equation}
    V^{\nu' \nu}_{\mu' \mu}(\kk, \kk', \qb)  = \displaystyle\sum_{\substack{a b \\ \sigma \sigma'}} V^{a b}_\qb \overline{\mathcal{U}}^{\mu'}_{\sigma a}(\kk+\qb) \mathcal{U}^\mu_{\sigma a}(\kk) \overline{\mathcal{W}}^{\nu'}_{\sigma' b}(\kk'-\qb) \mathcal{W}^{\nu}_{\sigma' b}(\kk'). \label{eq:Vindices}
\end{equation}
\end{widetext}
Here, a barred variable denotes complex conjugation.
The precise form of $V^{ab}_\qb$ depends on the stacking configuration between the two layers. For R-stacking, we find $V^{a b}_\qb =  U \delta^{a b} / N + V A^{a b}_\qb / N$, with 
\begin{subequations}
    \begin{align}
    A_\kk^{AB} &= 1 + e^{i \mathbf{b}_1\cdot \kk} + e^{i (\mathbf{b}_1 - \mathbf{b}_2)\cdot \kk}, \\
    A_\kk^{AC} &= 1 + e^{i \mathbf{b}_1\cdot \kk} + e^{i \mathbf{b}_2\cdot \kk}, \\
    A_\kk^{BC} &= 1 + e^{i \mathbf{b}_2\cdot \kk} + e^{i (\mathbf{b}_2-\mathbf{b}_1)\cdot \kk},
\end{align}
and $A_\kk^{aa}=0$, $A_\kk^{ba} = \overline{A}_\kk^{ab}$.
\end{subequations}

In the new basis~\eqref{eq:newbasis}, we write the exciton Eq.~\eqref{eq:exitonWavefunc} as
\begin{equation}
    \ket{X_\pbold} = x^\dagger_\pbold \ket{GS} = \displaystyle\sum_\kk \displaystyle\sum_{\lambda, \rho = 1}^6 \chi_\kk^{\lambda \rho} (\pbold) \gamma^\dagger_{\lambda \pbold-\kk} h^\dagger_{\rho \kk} \ket{GS}. \label{eq:excitonInGWC}
\end{equation}

The exciton wavefunction $\chi^{\lambda \rho}_{\kk}(\pbold)$ is obtained as the lowest-energy solution of the following equation
\begin{equation}
\begin{aligned}
    &\displaystyle\sum_{\nu'}\big[(\epsilon^\mu_{\pbold-\kk} + \Tilde{\epsilon}_{\kk}^{\nu})\delta_{\nu \nu'} + \vbar_{\nu \nu'}(\kk) \big] \nbar^\mu_{\pbold-\kk} \chi^{\mu \nu'}_{\kk}(\pbold)
     \\
     +  &\displaystyle\sum_{\substack{\kk' \\ \mu' \nu'}} V^{\nu \nu'}_{\mu \mu'}({\pbold-\kk', \kk', \kk'-\kk}) \nbar^{\mu'}_{\pbold-\kk'} \nbar^\mu_{\pbold-\kk} \chi^{\mu' \nu'}_{\kk'}(\pbold) \\
     &= E^X_\pbold \chi^{\mu \nu}_\kk(\pbold).  \label{eq:excitonSchroGWC}
\end{aligned}
\end{equation}
with $\vbar_{\nu \nu'}(\kk) = \sum_{\lambda \kk'} n^{\lambda}_{\kk'} V^{\nu \nu'}_{\lambda \lambda}(\kk', \kk, 0) $ and $\Tilde{\epsilon}_\kk^\lambda$ the hole dispersion in the new basis.
As remarked in the main text, the resulting exciton wavefunction does not necessarily represent an optically generated exciton. Instead, it describes the long-lived, lowest-energy exciton state to which optically created excitons relax. The relaxed, lowest-energy exciton state is relevant for photoluminescence and transport measurements, while an optical reflectance measurement would directly probe the optically excited exciton. Since reflectance measurements are difficult for interlayer excitons due to their small oscillator strength~\cite{riveraInterlayerValleyExcitons2018, torunInterlayerIntralayerExcitons2018}, we believe that the long-lived exciton states we focus on are experimentally more relevant. Nonetheless, assuming that one could selectively excite excitons at different energies, we could also target higher-energy exciton bound states in Eq.~\eqref{eq:excitonSchroGWC} and use them to study the polaron formation in our model.

\begin{figure*}
\begin{center}
\includegraphics[width=0.75\linewidth]{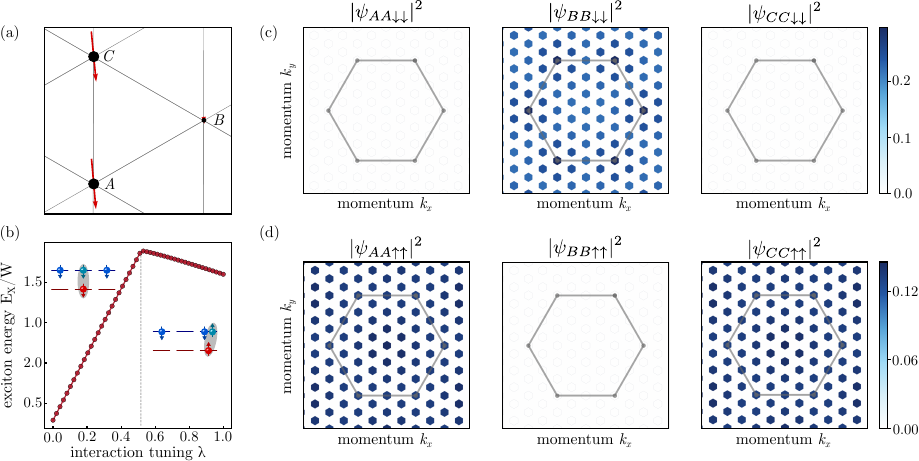}
\caption{\textbf{Exciton wavefunction in generalized Wigner crystals.} \textbf{(a)} Mean field solution for ferromagnetic generalized Wigner crystal at $\nu=2/3$, $U_{e-e}=16t$, $V_{e-e}=U_{e-e}/6$ and $X_{e-e} = 0.8t$. The two sublattices $A$ and $C$ are occupied, while $B$ is nearly empty. \textbf{(b)} Exciton energy as a function of a continuous tuning parameter $\lambda$, which linearly interpolates between $U_{e-h}=-15t$, $V_{e-h} = -10t$ ($\lambda=0$) and $U_{e-h}=-35t$, $V_{e-h} = 0$ ($\lambda=1$). For small $\lambda$, the exciton is formed on empty sites; for large $\lambda$, it forms on top of the crystal. \textbf{(c)} Sublattice and spin-resolved components of the exciton wavefunction $\psi_{ab, \kk}^{\sigma \sigma'}(\pbold=0)$, for $\lambda=0$. Here the exciton forms on an empty site, with the electron spin aligned with the spins of the crystal. \textbf{(d)} For $\lambda=1$, where the on-site attraction is strong, the exciton forms on top of the crystal. All other components of the wavefunction, which are not shown, vanish. System size $N= 3\times (6\times 6)$ and R-stacking.
}
\label{fig:X_GWC_WF}
\end{center}
\end{figure*}

After solving Eq.~\eqref{eq:excitonSchroGWC} in the Hartree-Fock basis, one can express the exciton wavefunction in the original basis again:
\begin{equation}
    \psi_{ab, \kk}^{\sigma \sigma'}(\pbold) = \displaystyle\sum_{\mu \nu} \chi^{\mu \nu}_{\kk }(\pbold) \mathcal{U}^\mu_{\sigma a}(\pbold -\kk) \mathcal{W}^\nu_{\sigma' b}(\kk).
\end{equation}
In the parameter regime discussed in the main text, with $|U_{e-h}|> U_{e-e}$, the exciton bound state is formed with an electron sitting atop the crystal; see Fig.~\ref{fig:X_GWC_WF} for $\nu=2/3$. The strength of the respective interactions will depend on material properties. For $|U_{e-h}| \lesssim U_{e-e}$, we find that a nearest-neighbor attractive interaction $V_{e-h}$ is required to form an exciton bound state, which now occupies an empty site next to the crystal. Note that such a configuration is only possible for generalized Wigner crystals, not for Mott states.

\textbf{\textit{Umklapp peaks.---}}
In the main text, we discussed the appearance of Umklapp peaks and how they can be used to distinguish between AFM and FM spin order based on symmetry arguments. Here, we provide a simple complementary description. The Umklapp peaks can be understood as the exciton scattering off with a particle-hole excitation of the crystal. A GWC at $\nu=2/3$ filling has two of every three sites filled, while the third one is empty; see inset in Fig.~\ref{fig:2}. For the AFM, one can have both particle-hole excitations from a filled to an empty site, and a particle-hole excitation from a filled site to another filled site (creating a doublon). 
The two Umklapp peaks correspond to the exciton scattering with these two different particle-hole excitations, where the Umklapp peak at higher energy corresponds to the scattering with a particle-hole excitation involving the formation of a doublon. Conversely, for the FM, the Pauli principle forbids a particle-hole excitation from a filled to another filled site of the crystal. This leaves only the Umklapp peak, where the exciton scatters with a particle-hole excitation from a filled site to an empty site.

\section{Determining the mass of exciton-polarons}
\label{app:determining}

\begin{figure*}
\begin{center}
\includegraphics[width=0.99\linewidth]{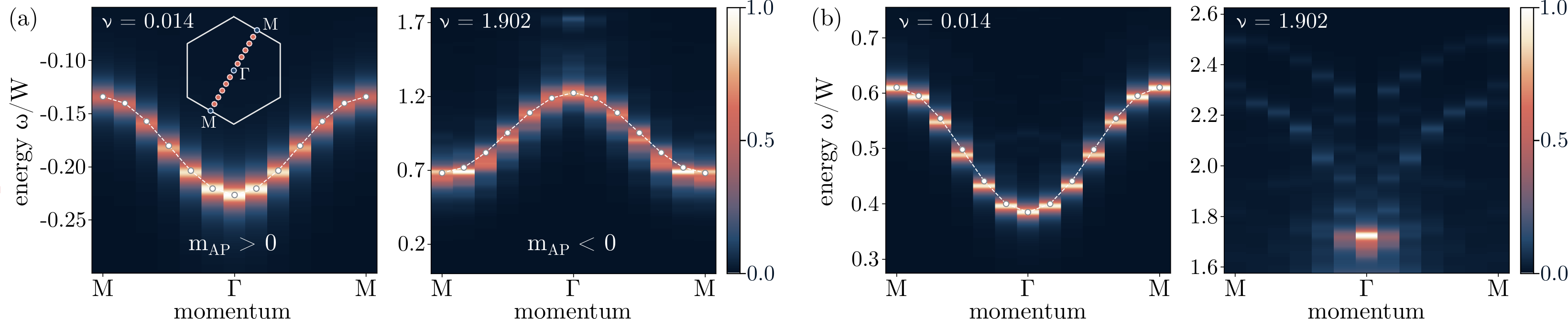}
\caption{\textbf{Momentum-dependent exciton-polaron spectral function (R-stacking).} \textbf{(a)} Left: Dispersion of attractive polaron in the low-density regime with a positive mass. Right: In the high-density regime, the attractive polaron has a negative effective mass. Note the vastly different energy scales in the low and high-density regime, with a much larger mobility for high densities. The white dotted line is the fit to the dispersion of a particle on a triangular lattice with nearest-neighbor hopping. The cut through the Brillouin zone is shown as an inset. \textbf{(b)} Dispersion for the RP branch. In the high-density regime, the RP is no longer a well-defined quasiparticle. We used $U_{e-h}=-12t$ for the electron-hole interaction; all other interactions are set to zero. System size of $N=12\times 12$ and R-stacking.
}
\label{fig:Supp_mass}
\end{center}
\end{figure*}

\begin{figure*}
\begin{center}
\includegraphics[width=0.99\linewidth]{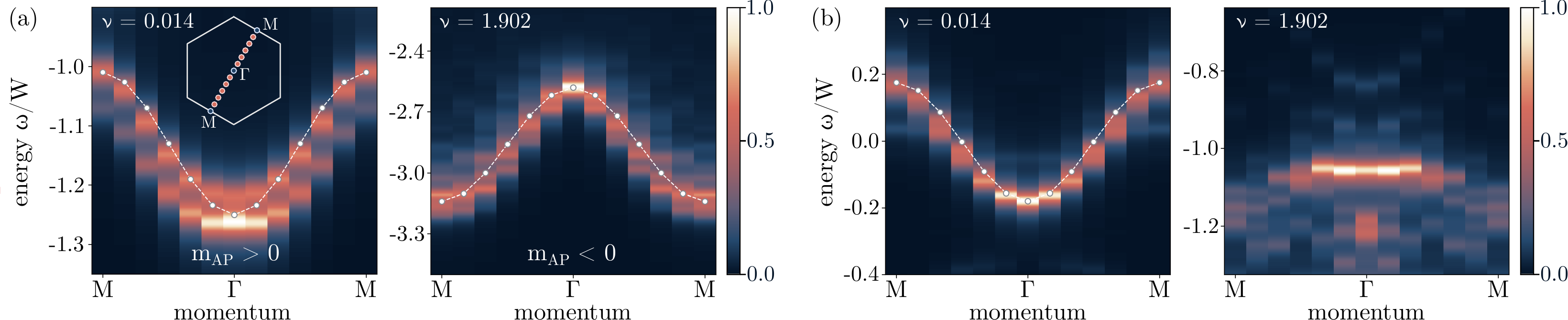}
\caption{\textbf{Momentum-dependent exciton-polaron spectral function (H-stacking).} \textbf{(a)} Left: Dispersion of attractive polaron in the low-density regime with a positive mass. Right: In the high-density regime, the attractive polaron has a negative effective mass. Note the vastly different energy scales in the low and high-density regime, with a much larger mobility for high densities. The white dotted line is the fit to the dispersion of a particle on a triangular lattice with nearest-neighbor hopping. The cut through the Brillouin zone is shown as an inset. \textbf{(b)} Dispersion for the RP branch. In the high-density regime, the RP is no longer a well-defined quasiparticle. We used $U_{e-h}=-15t$ for the electron-hole interaction; all other interactions are set to zero. System size of $N=12\times 12$ and H-stacking.
}
\label{fig:Supp_mass2}
\end{center}
\end{figure*}

\begin{figure*}
\begin{center}
\includegraphics[width=0.75\linewidth]{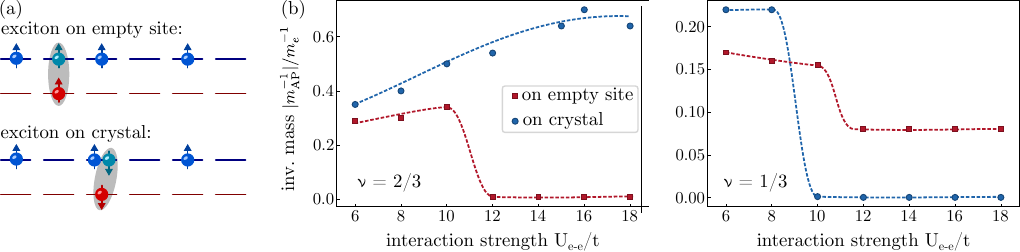}
\caption{\textbf{Exciton mass renormalization in generalized Wigner crystals.} \textbf{(a)} If the on-site attractive interaction is small, $|U_{e-h}| \lesssim U_{e-e}$, and $V_{e-h}$ sufficiently large, the exciton forms on an empty site (top). In contrast, for large $|U_{e-h}| > U_{e-e}$, doubly occupancy is favored, leading to an exciton formed atop the crystal (bottom). \textbf{(b)} Exciton-polaron mobility in generalized Wigner crystals for $\nu=2/3$ (left) and $\nu=1/3$ (right) is strongly affected by whether the exciton is on an empty site or atop the crystal. For the empty site exciton, we use $U_{e-h} = -15t$ and $V_{e-h}= -10t$, for the exciton atop the crystal $U_{e-h} = -35t$ and $V_{e-h}= 0$. In the $\nu=2/3$ crystal $V_{e-e}= U_{e-e}/6$ and in the $\nu=1/3$ crystal $V_{e-e}= U_{e-e}/3$. System size $N=3\times(6\times6)$ and R-stacking.
}
\label{fig:X_compare}
\end{center}
\end{figure*}

We extract the effective mass of the attractive polaron from the momentum-dependent exciton spectral function $\mathcal{A}_X(\kk, \omega) =  -\frac{1}{\pi} \Im \mathcal{G}_X(\kk, \omega)$. 

We define the mass $m$ of a particle with dispersion $\varepsilon_\kk$ as $m^{-1} = \partial^2 \varepsilon_\kk/ \partial \kk^2 |_{\kk=0}$. For a particle hopping on a triangular lattice, the mass is given by $m=3/W$, where $W$ is the bandwidth of $\varepsilon_\kk$. We use this relation to determine the mass of the AP by calculating its bandwidth.

Concretely, we calculate the bandwidth of the attractive polaron in units of the bare electron bandwidth $W=9t$. For a particle hopping on a triangular lattice, its mass is related to the bandwidth though $m=3/W$, which follows from $m^{-1} = \partial^2 \varepsilon_\kk/ \partial \kk^2 |_{\kk=0}$ and $\varepsilon_\kk$ The sign of $t_{AP}$ is determined by the curvature of the dispersion at the $\Gamma$ point. The mass of the attractive polaron is inversely proportional to the hopping $t_{AP}$. For low-electron densities, the mass is positive, while in the high-density regime, it is negative; see Fig.~\ref{fig:Supp_mass}(a). In the low-density regime, the RP is also a quasiparticle with a well-defined dispersion. At the same electron densities, we find that the RP mass is smaller than the AP mass; compare Fig.~\ref{fig:Supp_mass}(a) and (b). In the high-density regime, the RP no longer has a well-defined dispersion, but rather exhibits a spectral weight peak at zero momentum. This indicates that the RP is no longer a well-defined quasiparticle. 

For Fig.~\ref{fig:3}(b) in the main text, we used $U_{e-h}=-12t$ and $V_{e-h}=-3t$ with a system size of $N=9\times9$ in the low- and high-density regime, while we used $U_{e-h}=-35t$ and a system size of $N=3\times(6\times6)$ for the correlated insulators at the fillings $\nu \in \{1/3, 2/3, 1\}$. For lower fillings, stronger long-range interactions $V_{e-e}$ are required to stabilize the crystal order. For $\nu=1$, $\nu=2/3$ and $\nu=1/3$, we use $V_{e-e}=0$, $V_{e-e} = U_{e-h}/6$ and $V_{e-e} = U_{e-h}/3$ respectively. We find that exciton mobility is strongly affected by the different exciton wavefunctions. As discussed in the previous section, the exciton forms on top of the crystal for large $|U_{e-h}|$, while it sits on an empty site for smaller $|U_{e-h}|$ and larger $|V_{e-h}|$; see Figs.~\ref{fig:X_GWC_WF} and~\ref{fig:X_compare}(a). In the latter case, for $\nu=1/3$, the exciton-polaron can hop between empty sites, leading to enhanced mobility compared to the localized exciton atop the crystal. The situation is reversed for the $\nu=2/3$ state since empty sites are not connected to each other; see Fig.~\ref{fig:X_compare}(b).

\section{Chevy approximation} \label{sec:Chevy}

Using the Chevy approximation, we approach the many-body problem of a single exciton immersed in an electronic ground state. We calculate the exciton propagator $\mathcal{G}_X(\pbold, \omega)$, defined in Eq.~\eqref{eq:Xprop}, by rewriting it as 

\begin{equation}
    \mathcal{G}_X(\kk, \omega) = \ip{X_\kk}{\Psi_\kk(\omega)},
\end{equation}
with
\begin{equation}
    \ket{\Psi_\kk(\omega)} = \frac{1}{\omega + i \eta - H} \ket{X_\kk}.
\end{equation}
We obtain the state $\ket{\Psi_\kk(\omega)}$ by iteratively solving the linear equation
\begin{equation}
    {(\omega + i \eta - H)} \ket{\Psi_\kk(\omega)} = \ket{X_\kk} \label{eq:itereq}
\end{equation}
for each $\omega$ and some finite regularization $\eta$. To make this computationally feasible, we use the Chevy approximation, which restricts the Hilbert space to states with a single exciton and an exciton with a particle-hole excitation of the electronic ground state. Concretely, we define the following basis states: 

\begin{align*}
    \ket{n=0} &= \ket{X_\pbold} \equiv x^\dagger_\pbold \ket{GS} \\
    \ket{n>0} &= \ket{C^\pbold_{\kk \alpha \qb \beta}} \equiv x^\dagger_{\pbold+\qb-\kk} c^\dagger_{\alpha \kk} c_{\beta \qb} \ket{GS} \label{eq:Chevybasis_supp} \numberthis
\end{align*}
and compute the matrix elements of the Hamiltonian in this basis:
\begin{equation}
    H_{nm} = \bra{n} H \ket{m}. \label{eq:Hammatrix}
\end{equation}
Note that we treat the electron-electron interaction only on a mean-field level. Consequently, its only effect is to renormalize the electron dispersion.
In the conventional Chevy ansatz, where the exciton is treated as a bosonic point-like particle, the Chevy basis Eq.~\eqref{eq:Chevybasis_supp} is orthogonal. But since we take the internal structure of the exciton into account, this is no longer the case
\begin{equation}
    b_{n m} := \ip{n}{m} \neq \delta_{n m}.
\end{equation}

\begin{widetext}

Projected to the Chevy basis, Eq.~\eqref{eq:itereq} reduces to the following matrix equation
\begin{equation}
    \displaystyle\sum_m \big[(\omega + i\eta)b_{nm} - H_{nm} \big] \Psi^m_\kk (\omega) = X^n_{\kk} \qq{with} \Psi^n_\kk (\omega) = \ip{n}{\Psi_\kk(\omega)} \qq{and} X^n_\kk = \ip{n}{X_\kk}.
\end{equation}

The matrix elements of the Hamiltonian Eq.~\eqref{eq:Hammatrix} can be computed explicitly
\begin{subequations} \label{eq:projection}
\begin{flalign*}
    \mel{X_\pbold}{H}{X_\pbold} &= \displaystyle\sum_\kk |\psi_{\pbold-\kk}(\pbold)|^2 \nbar_{\uparrow \kk} ( \eps^e_\kk + \eps^h_{\pbold-\kk} + \nu V_{\qb=0} ) + \frac{1}{N} \displaystyle\sum_{\kk \qb} V_{\qb} \overline{\psi}_{\pbold-\kk-\qb}(\pbold)\psi_{\pbold-\kk}(\pbold)\nbar_{\uparrow \kk} \nbar_{\uparrow \kk+\qb}, &\numberthis \label{eq:XHX_1}
\end{flalign*}
\begin{flalign*}
    \mel{C^\pbold_{\kk \alpha \qb \beta}}{H}{X_\pbold} &= -\overline{\psi}_{\pbold-\kk}(\pbold)\psi_{\pbold-\kk}(\pbold+\qb-\kk) ( \eps^e_\kk + \eps^h_{\pbold-\kk}  + \nu V_{\pbold=0})\delta_{\alpha \uparrow} \delta_{\beta \uparrow}& \\
    &\quad- \frac{1}{N} \displaystyle\sum_{\qb'} \bigg[ V_{\qb'} \overline{\psi}_{\pbold-\kk-\qb'}(\pbold)\psi_{\pbold-\kk}(\pbold+\qb-\kk)\nbar_{\uparrow\kk+\qb'} + V_{\qb'} \overline{\psi}_{\pbold-\kk}(\pbold)\psi_{\pbold-\kk+\qb'}(\pbold+\qb-\kk)\nbar_{\uparrow \qb-\qb'} & \\
    &\quad- V_{\qb-\kk} \overline{\psi}_{\qb'-\qb+\kk}(\pbold)\psi_{\qb'}(\pbold+\qb-\kk)\nbar_{\uparrow \pbold+\qb-\kk-\qb'}\bigg]\delta_{\alpha \uparrow} \delta_{\beta \uparrow}, &\numberthis
\end{flalign*}
\begin{flalign*}
    \mel{C^\pbold_{\kk \alpha \qb \beta}}{H}{C^\pbold_{\kk' \alpha' \qb' \beta'}} &= \Gamma^\pbold_1(\kk, \alpha, \qb, \beta) \delta_{\alpha \alpha'}\delta_{\beta \beta'}\delta_{\kk \kk'}\delta_{\qb \qb'} + \Gamma^\pbold_2(\kk, \alpha, \qb, \beta; \qb', \beta') \delta_{\alpha \alpha'}\delta_{\kk \kk'} & \\
    &\quad+ \Gamma^\pbold_3(\kk, \alpha, \qb, \beta; \kk', \alpha') \delta_{\beta \beta'}\delta_{\qb \qb'} + \Gamma^\pbold_4(\kk, \alpha, \qb, \beta; \kk', \alpha', \qb', \beta'), & \numberthis
\end{flalign*}
    \end{subequations}
with 
{
\begin{subequations} \label{eq:Gammas}
\begin{flalign*}
    \Gamma^\pbold_1(\kk, \alpha, \qb, \beta) &= \displaystyle\sum_{\pbold'} |\psi_{\pbold'}(\pbold+\qb-\kk)|^2 \nbar_{\uparrow \pbold + \qb -\kk -\pbold'} (\eps^e_{\pbold + \qb -\kk - \pbold'} + \eps^e_{\kk} - \eps^e_{\qb} + \eps^h_{\pbold'} + \nu V_{\pbold=0}) &
        \\
        &\quad+ \frac{1}{N} \displaystyle\sum_{\pbold' \qb'} V_{\qb'} \overline{\psi}_{\pbold'-\qb'}(\pbold+\qb-\kk) \psi_{\pbold'}(\pbold+\qb-\kk)\nbar_{\uparrow\pbold+\qb-\kk-\pbold'} \nbar_{\uparrow \pbold+\qb-\kk-\pbold'+\qb'}, &\numberthis
    \end{flalign*}

\begin{flalign*}
    \Gamma^\pbold_2(\kk, \alpha, \qb, \beta; \qb', \beta') &= \overline{\psi}_{\pbold-\kk}(\pbold+\qb-\kk)\psi_{\pbold-\kk}(\pbold+\qb'-\kk) (\eps^e_{\kk} + \eps^h_{\pbold-\kk} + \nu V_{\pbold=0}) \delta_{\beta \uparrow} \delta_{\beta' \uparrow} &
    \\
    &\quad+ \frac{1}{N} \displaystyle\sum_{\pbold'} \Bigg[V_{\pbold'} \overline{\psi}_{\pbold-\kk-\pbold'}(\pbold+\qb-\kk) \psi_{\pbold-\kk}(\pbold+\qb'-\kk)\nbar_{\uparrow\qb+\pbold'}\delta_{\beta \uparrow} \delta_{\beta' \uparrow}  & \\
    &\quad + V_{\pbold'} \overline{\psi}_{\pbold-\kk}(\pbold+\qb-\kk) \psi_{\pbold-\kk+\pbold'}(\pbold+\qb'-\kk)\nbar_{\uparrow\qb'-\pbold'}\delta_{\beta \uparrow} \delta_{\beta' \uparrow} & \\
    &\quad - V_{\qb'-\qb} \overline{\psi}_{\pbold'-\qb'+\qb}(\pbold+\qb-\kk) \psi_{\pbold'}(\pbold+\qb'-\kk)\nbar_{\uparrow\pbold-\kk-\pbold'-\qb'} \delta_{\beta \beta'}\Bigg], & \numberthis   
\end{flalign*}

\begin{flalign*}
    \Gamma^\pbold_3(\kk, \alpha, \qb, \beta; \kk', \alpha') &=-\overline{\psi}_{\pbold+\qb-\kk-\kk'}(\pbold+\qb-\kk)\psi_{\pbold+\qb-\kk-\kk'}(\pbold+\qb-\kk') (\eps^e_{\kk} + \eps^e_{\kk'} - \eps^e_{\qb} + \eps^h_{\pbold+\qb-\kk-\kk'} + \nu V_{\pbold=0}) \delta_{\alpha \uparrow} \delta_{\alpha' \uparrow} &
    \\
    &\quad- \frac{1}{N} \displaystyle\sum_{\pbold'} \Bigg[V_{\pbold'} \overline{\psi}_{\pbold+\qb-\kk-\kk'}(\pbold+\qb-\kk) \psi_{\pbold+\qb-\kk-\kk'+\pbold'}(\pbold+\qb-\kk')\nbar_{\uparrow\kk-\pbold'}\delta_{\alpha \uparrow} \delta_{\alpha' \uparrow} &   \\
    &\quad + V_{\pbold'} \overline{\psi}_{\pbold+\qb-\kk-\kk'-\pbold'}(\pbold+\qb-\kk) \psi_{\pbold+\qb-\kk-\kk'}(\pbold+\qb-\kk')\nbar_{\uparrow\kk'+\pbold'}\delta_{\alpha \uparrow} \delta_{\alpha' \uparrow} & \\
    &\quad - V_{\kk-\kk'} \overline{\psi}_{\pbold'-\kk+\kk'}(\pbold+\qb-\kk) \psi_{\pbold'}(\pbold+\qb-\kk')\nbar_{\uparrow\pbold+\qb-\kk'-\pbold'} \delta_{\alpha \alpha'}\Bigg], & \numberthis
\end{flalign*}

\begin{flalign*}
    \Gamma^\pbold_4(\kk, \alpha, \qb, \beta; \kk', \alpha', \qb', \beta') &= -V_{\kk-\qb} \overline{\psi}_{\pbold+\qb-\kk-\kk'}(\pbold+\qb-\kk) \psi_{\pbold-\kk'}(\pbold+\qb'-\kk') \delta_{\alpha \beta} \delta_{\alpha' \uparrow} \delta_{\beta' \uparrow} &\\
    &\quad - V_{\qb'-\kk'} \overline{\psi}_{\pbold-\kk}(\pbold+\qb-\kk) \psi_{\pbold-\kk-\kk'-\qb'}(\pbold+\qb'-\kk') \delta_{\alpha' \beta'} \delta_{\alpha \uparrow} \delta_{\beta \uparrow} & \\
    &\quad + V_{\kk-\kk'} \overline{\psi}_{\pbold-\kk}(\pbold+\qb-\kk) \psi_{\pbold-\kk'}(\pbold+\qb'-\kk') \delta_{\alpha \alpha'} \delta_{\beta \uparrow} \delta_{\beta' \uparrow} & \\
    &\quad + V_{\qb'-\qb} \overline{\psi}_{\pbold+\qb-\kk-\kk'}(\pbold+\qb-\kk) \psi_{\pbold+\qb'-\kk-\kk'}(\pbold+\qb'-\kk') \delta_{\beta \beta'} \delta_{\alpha \uparrow} \delta_{\alpha' \uparrow}. & \numberthis
\end{flalign*}
    
\end{subequations}
}

For correlated insulators, the Chevy calculation can be carried out analogously: Expressed in terms of the Hartree-Fock quasiparticles defined in Eq.~\eqref{eq:newbasis}, the Chevy basis becomes \begin{subequations}\label{eq:ChevybasisGWC}
\begin{align}
    \ket{n=0} &= \ket{X_\pbold} \equiv x^\dagger_\pbold \ket{GS} \label{eq:ChevybasisGWC_a}, \\
    \ket{n>0} &= \ket{C^\pbold_{\kk \mu \qb \nu}} \equiv x^\dagger_{\pbold+\qb-\kk} \gamma^\dagger_{\mu \kk} \gamma_{\nu \qb} \ket{GS} \label{eq:ChevybasisGWC_b}. 
\end{align}
\end{subequations}
and the matrix elements of the Hamiltonian are given by
\begin{subequations}\label{eq:projection_GWC}
\begin{flalign*}
    \mel{X_\pbold}{H}{X_\pbold} &= \displaystyle\sum_{\substack{\kk \\ \mu \nu}} \overline{\chi}^{\mu \nu'}_{\pbold-\kk}(\pbold) \chi^{\mu \nu}_{\pbold-\kk}(\pbold)  \big[(\epsilon^\mu_{\kk} + \Tilde{\epsilon}_{\pbold -\kk}^{\nu})\delta_{\nu \nu'} + \vbar_{\nu \nu'}(\pbold -\kk) \big] \nbar^\mu_{\kk}& \\
    &\quad +\frac{1}{N} \displaystyle\sum_{\kk \qb } \displaystyle\sum_{ \mu \nu \mu' \nu'} V_{\mu' \mu}^{\nu' \nu}(\kk, \pbold-\kk, \qb) \overline{\chi}^{\mu' \nu'}_{\pbold-\kk-\qb}(\pbold)\chi^{\mu \nu}_{\pbold-\kk}(\pbold)\nbar^\mu_{\kk} \nbar^{\mu'}_{\uparrow \kk+\qb}, &\numberthis
\end{flalign*}
\begin{flalign*}
    \mel{C^\pbold_{\kk \alpha \qb \beta}}{H}{X_\pbold} &= -\displaystyle\sum_{\nu \nu'}\overline{\chi}^{\alpha \nu'}_{\pbold-\kk}(\pbold) \chi^{\beta \nu}_{\pbold-\kk}(\pbold+\qb-\kk) \big[( \epsilon^\alpha_\kk + \Tilde{\epsilon}^\nu_{\pbold-\kk})\delta_{\nu \nu'}  + \vbar_{\nu' \nu}(\pbold-\kk) \big]& \\
    &\quad- \frac{1}{N} \displaystyle\sum_{\substack{\qb' \\ \mu \nu \nu'}} \bigg[ V^{\nu' \nu}_{\mu \alpha}(\kk, \pbold-\kk, \qb') \bar{\chi}^{\mu \nu'}_{\pbold-\kk-\qb'}(\pbold) \chi^{\beta \nu}_{\pbold-\kk}(\pbold+\qb-\kk)\nbar^{\mu}_{\kk+\qb'}  &\\ 
    &\quad + \overline{V}^{\nu \nu'}_{\mu \beta}(\qb, \pbold-\kk, -\qb') \overline{\chi}^{\mu \nu}_{\pbold-\kk}(\pbold) \chi^{\mu \nu}_{\pbold-\kk+\qb'}(\pbold+\qb-\kk)\nbar^{\mu}_{\qb-\qb'} & \\
    &\quad - V^{\nu \nu'}_{\beta \alpha}(\kk, \qb', \qb-\kk) \overline{\chi}^{\mu \nu'}_{\qb'-\qb+\kk}(\pbold) \chi^{\mu \nu'}_{\qb'}(\pbold+\qb-\kk)\nbar^\mu_{\pbold+\qb-\kk-\qb'}\bigg], &\numberthis
\end{flalign*}
\begin{flalign*}
    \mel{C^\pbold_{\kk \alpha \qb \beta}}{H}{C^\pbold_{\kk' \alpha' \qb' \beta'}} &= \Gamma^\pbold_1(\kk, \alpha, \qb, \beta) \delta_{\alpha \alpha'}\delta_{\beta \beta'}\delta_{\kk \kk'}\delta_{\qb \qb'} + \Gamma^\pbold_2(\kk, \alpha, \qb, \beta; \qb', \beta') \delta_{\alpha \alpha'}\delta_{\kk \kk'} & \\
    &\quad+ \Gamma^\pbold_3(\kk, \alpha, \qb, \beta; \kk', \alpha') \delta_{\beta \beta'}\delta_{\qb \qb'} + \Gamma^\pbold_4(\kk, \alpha, \qb, \beta; \kk', \alpha', \qb', \beta'), & \numberthis
\end{flalign*}
    \end{subequations}
with
{
\begin{subequations}\label{eq:Gammas_GWC}
\begin{flalign*}
    \Gamma^\pbold_1(\kk, \alpha, \qb, \beta) &= \displaystyle\sum_{\substack{\pbold'\\ \mu \nu \nu'}} \overline{\chi}^{\mu \nu'}_{\pbold'}(\pbold+\qb-\kk) \chi^{\mu \nu}_{\pbold'}(\pbold+\qb-\kk) \nbar^{\mu}_{\pbold+\qb-\kk-\pbold'} \big[(\epsilon^\mu_{\pbold+\qb-\kk-\pbold'} + \Tilde{\epsilon}_{\pbold'}^{\nu} + \epsilon^\alpha_\kk-\epsilon^\beta_{\qb})\delta_{\nu \nu'} + \vbar_{\nu' \nu}(\pbold') \big]  &
        \\
        &\quad+ \frac{1}{N} \displaystyle\sum_{\substack{\pbold' \qb' \\ \mu \mu' \nu \nu'}} V^{\nu' \nu}_{\mu' \mu}(\pbold+\qb-\kk-\pbold', \pbold', \qb') \overline{\chi}^{\mu' \nu'}_{\pbold'-\qb'}(\pbold+\qb-\kk) \chi^{\mu \nu}_{\pbold'}(\pbold+\qb-\kk)\nbar^{\mu}_{\pbold+\qb-\kk-\pbold'} \nbar^{\mu'}_{ \pbold+\qb-\kk-\pbold'+\qb'} , &\numberthis
    \end{flalign*}

\begin{flalign*}
    \Gamma^\pbold_2(\kk, \alpha, \qb, \beta; \qb', \beta') &=\displaystyle\sum_{\nu \nu'} \overline{\chi}^{\beta \nu'}_{\pbold-\kk}(\pbold+\qb-\kk) \chi^{\beta' \nu}_{\pbold-\kk}(\pbold+\qb'-\kk) \big[ (\epsilon^\alpha_{\kk} + \Tilde{\epsilon}^\nu_{\pbold-\kk})\delta_{\nu \nu'} + \vbar_{\nu' \nu}(\pbold-\kk) \big]  &
    \\
    &\quad+ \frac{1}{N} \displaystyle\sum_{\substack{\pbold' \\ \mu \nu \nu'}} \Bigg[V^{\nu' \nu}_{\mu \beta}(\qb, \pbold-\kk, \pbold') \overline{\chi}^{\mu \nu'}_{\pbold-\kk-\pbold'}(\pbold+\qb-\kk) \chi^{\beta' \nu}_{\pbold-\kk}(\pbold+\qb'-\kk)\nbar^{\mu}_{\qb+\pbold'}  & \\
    &\quad + \overline{V}^{\nu \nu'}_{\mu \beta'}(\qb', \pbold-\kk, -\pbold') \overline{\chi}^{\beta \nu'}_{\pbold-\kk}(\pbold+\qb-\kk) \chi^{\mu \nu}_{\pbold-\kk+\pbold'}(\pbold+\qb'-\kk)\nbar^{\mu}_{\qb'-\pbold'} & \\
    &\quad - V^{\nu' \nu}_{\beta' \beta}(\qb, \pbold', \qb'-\qb) \overline{\chi}^{\mu \nu'}_{\pbold'-\qb'+\qb}(\pbold+\qb-\kk) \chi^{\mu \nu}_{\pbold'}(\pbold+\qb'-\kk)\nbar^{\mu}_{\pbold-\kk-\pbold'-\qb'} \Bigg], & \numberthis   
\end{flalign*}

\begin{flalign*}
    \Gamma^\pbold_3(\kk, \alpha, \qb, \beta; \kk', \alpha') &=-\displaystyle\sum_{\nu \nu'}\overline{\chi}^{\alpha' \nu'}_{\pbold+\qb-\kk-\kk'}(\pbold+\qb-\kk)\chi^{\alpha \nu}_{\pbold+\qb-\kk-\kk'}(\pbold+\qb-\kk') &\\ 
    &\quad \times \big[ (\epsilon^\alpha_{\kk} + \epsilon^{\alpha'}_{\kk'} - \epsilon^\beta_{\qb} + \Tilde{\eps}^{\nu}_{\pbold+\qb-\kk-\kk'})\delta_{\nu \nu'} + \vbar_{\nu' \nu}(\pbold + \qb-\kk-\kk') \big] &
    \\
    &\quad- \frac{1}{N} \displaystyle\sum_{\substack{\pbold' \\ \mu \nu \nu'}} \Bigg[\overline{V}^{\nu \nu'}_{\mu \alpha}(\kk, \pbold+\qb-\kk-\kk', -\pbold') \overline{\chi}^{\alpha' \nu'}_{\pbold+\qb-\kk-\kk'}(\pbold+\qb-\kk) \chi^{\mu \nu}_{\pbold+\qb-\kk-\kk'+\pbold'}(\pbold+\qb-\kk')\nbar^{\mu}_{\kk-\pbold'}&   \\
    &\quad + V^{\nu' \nu}_{\mu \alpha'}(\kk', \pbold+\qb-\kk-\kk', \pbold') \overline{\chi}^{\mu \nu'}_{\pbold+\qb-\kk-\kk'-\pbold'}(\pbold+\qb-\kk) \chi^{\alpha \nu}_{\pbold+\qb-\kk-\kk'}(\pbold+\qb-\kk')\nbar^{\mu}_{\kk'+\pbold'}& \\
    &\quad - V^{\nu' \nu}_{\alpha \alpha'}(\kk', \qb', \kk-\kk') \overline{\chi}^{\mu \nu'}_{\pbold'-\kk+\kk'}(\pbold+\qb-\kk) \chi^{\mu \nu}_{\pbold'}(\pbold+\qb-\kk')\nbar^{\mu}_{\pbold+\qb-\kk'-\pbold'} \Bigg], & \numberthis
\end{flalign*}

\begin{flalign*}
    \Gamma^\pbold_4(\kk, \alpha, \qb, \beta; \kk', \alpha', \qb', \beta') &= - \displaystyle\sum_{\nu \nu'} \Bigg[ V^{\nu' \nu}_{\alpha \beta}(\qb, \pbold-\kk', \kk-\qb) \overline{\chi}^{\alpha' \nu'}_{\pbold+\qb-\kk-\kk'}(\pbold+\qb-\kk) \chi^{\beta' \nu}_{\pbold-\kk'}(\pbold+\qb'-\kk')  &\\
    &\quad + \overline{V}^{\nu \nu'}_{\alpha' \beta'}(\qb', \pbold-\kk, \kk'-\qb') \overline{\chi}^{\beta \nu'}_{\pbold-\kk}(\pbold+\qb-\kk) \chi^{\alpha \nu}_{\pbold-\kk-\kk'-\qb'}(\pbold+\qb'-\kk')  & \\
    &\quad - V^{\nu' \nu}_{\alpha \alpha'}(\kk', \pbold-\kk', \kk-\kk') \overline{\chi}^{\beta \nu'}_{\pbold-\kk}(\pbold+\qb-\kk) \chi^{\beta' \nu}_{\pbold-\kk'}(\pbold+\qb'-\kk') & \\
    &\quad - V^{\nu' \nu}_{\beta' \beta}(\qb, \pbold+\qb'-\kk-\kk', \qb'-\qb) \overline{\chi}^{\alpha' \nu'}_{\pbold+\qb-\kk-\kk'}(\pbold+\qb-\kk) \chi^{\alpha \nu}_{\pbold+\qb'-\kk-\kk'}(\pbold+\qb'-\kk') \Bigg]. & \numberthis
\end{flalign*}
    
\end{subequations}
}

\section{Diagrammatic representation of Chevy calculation}
\label{app:diagrams}

The matrix elements of the Hamiltonian with respect to the Chevy basis, given in Eqs.~\eqref{eq:projection}, \eqref{eq:Gammas},~\eqref{eq:projection_GWC}, and~\eqref{eq:Gammas_GWC}, can be intuitively interpreted using diagrammatic notation. To that end, we represent the exciton wavefunction diagrammatically as follows
\begin{equation}
    \psi_\kk(\pbold) = \raisebox{-3.8ex}{\includegraphics[height=9ex]{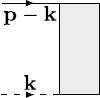}} \qq{and} \chi_\kk^{\mu \nu}(\pbold) = \raisebox{-3.8ex}{\includegraphics[height=9.5ex]{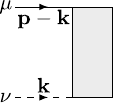}}
\end{equation}
where a solid line with an arrow corresponds to an electron and a dashed line with an arrow to a hole. The vertex for the attractive electron-hole interaction is written as
\begin{equation}
    V_\qb =  \raisebox{-4.8ex}{\includegraphics[height=11ex]{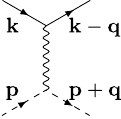}} \qq{and} V_{\mu' \mu}^{\nu' \nu}(\kk, \pbold, \qb) = \raisebox{-4.8ex}{\includegraphics[height=11ex]{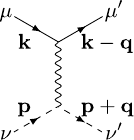}},
\end{equation}
for the simple electron-hole interaction and the interaction between holes and Hartree-Fock quasiparticles of correlated insulators~\eqref{eq:Vindices}, respectively. 
When projecting to the Chevy basis Eq.~\eqref{eq:Chevybasis_supp}, we highlight external momenta, belonging to the particle-hole excitations of the electronic ground state, with a colored node. Blue represents particle momenta and red momenta of the hole excitation. We use generalized momentum indices, including the spin/band degrees of freedom: $q = (\qb, \mu)$. A $\delta$-function is diagrammatically represented as follows: $\delta_{k, k'} = k \, \raisebox{-0.2ex}{\includegraphics[width=7ex]{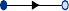}} \, k'$  and $\delta_{q, q'} = q \, \raisebox{-0.2ex}{\includegraphics[width=7ex]{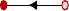}} \, q'$.
We present the diagrams contributing to Eqs.~\eqref{eq:projection} and~\eqref{eq:Gammas}. For simplicity, we do not explicitly include kinetic and occupation factors $\theta_\kk = 1 - n_\kk$.
The projection of the Hamiltonian to the space with just a single exciton Eq.~\eqref{eq:XHX_1} has no external momentum indices and is given by three diagrams:
\begin{subequations}
\begin{flalign}
        \mel{X_\pbold}{H}{X_\pbold} \sim \raisebox{-3.2ex}{\includegraphics[height=7.1ex]{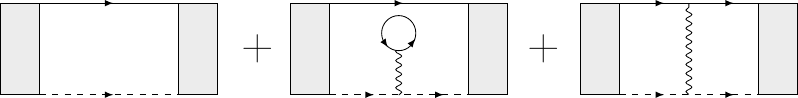}} . \label{eq:diagrams_1}
\end{flalign}
For the remaining terms in Eqs.~\eqref{eq:projection} and~\eqref{eq:Gammas}, we only show the diagrams with an interaction vertex. Note that additionally, there would also be the same diagram without interaction vertex, equivalent to the first diagram in Eq.~\eqref{eq:diagrams_1}, for all terms except for $\Gamma_4$.
\begin{flalign}
        \mel{C^\pbold_{k q}}{H}{X_\pbold} \sim \raisebox{-8ex}{\includegraphics[height=14ex]{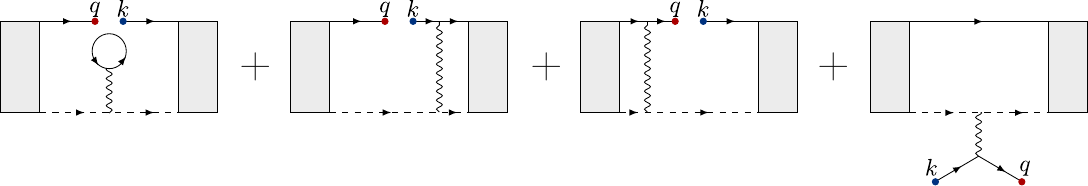}} 
\end{flalign}
\end{subequations}

\begin{subequations}
\begin{flalign}
        &\Gamma_1^\pbold(k, q) \delta_{k, k'} \delta_{q, q'} \sim \raisebox{-4.2ex}{\includegraphics[height=9ex]{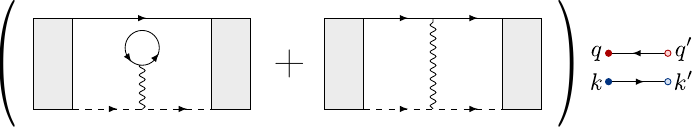}} &
\end{flalign}
\begin{flalign}
    &\Gamma_2^\pbold(k, q, q') \delta_{k, k'} \sim \raisebox{-8ex}{\includegraphics[height=14ex]{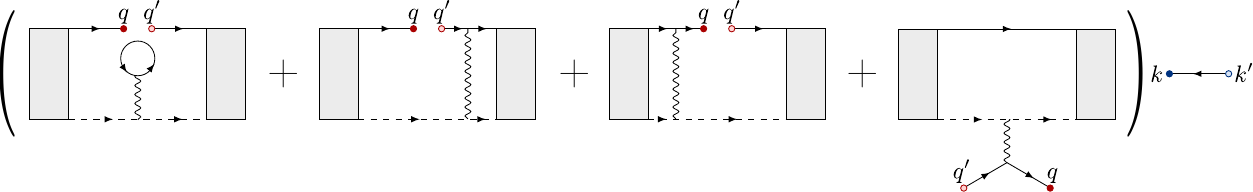}} &
\end{flalign}
\begin{flalign}
    &\Gamma_3^\pbold(k, q, k') \delta_{q, q'} \sim \raisebox{-8ex}{\includegraphics[height=14ex]{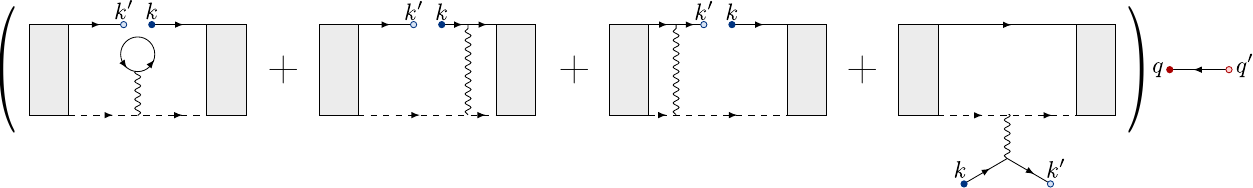}} &
\end{flalign}
\begin{flalign}
    &\Gamma_4^\pbold(k, q, k', q') \sim \raisebox{-8ex}{\includegraphics[height=14ex]{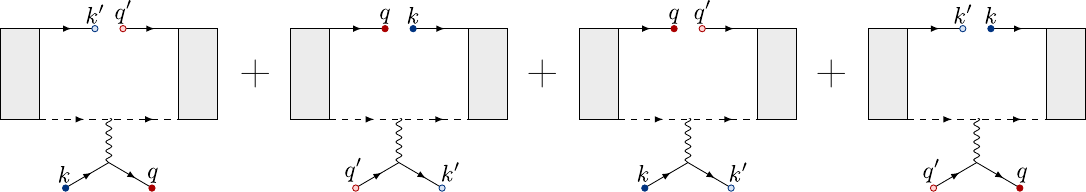}} &
\end{flalign}
\end{subequations}

\end{widetext}

\bibliography{bibfile}

\end{document}